\documentclass[preprint2]{aastex}

\usepackage{amssymb,lscape}
\usepackage[utf8]{inputenc}

\newcommand\gr{\mbox{\,g}}
\newcommand\cm{\mbox{\,cm}}

\newcommand\pc{\mbox{\,pc}}

\newcommand\Myr{\mbox{\,Myr}}

\newcommand\kms{\mbox{\,km s}^{-1}}
\newcommand\nsf{n_{\rm sf}}

\newcommand\eff{{\rm eff}}
\newcommand\Msun{M_\odot}

\newcommand\psc{\mbox{\,cm}^{-2}}
\newcommand\pcc{\mbox{\,cm}^{-3}}

\newcommand\dif{\mathrm{d}}

\newcommand\vlos{v_{\mbox{\scriptsize los}}}

\slugcomment{Submitted to ApJ}

\shorttitle{Filaments in MC simulations}
\shortauthors{G\'omez \& V\'azquez-Semadeni}

\begin{document}

\title{Filaments in simulations of molecular cloud formation}

\author{Gilberto C. G\'omez \& Enrique V\'azquez-Semadeni}
\affil{Centro de Radioastronom\1a y Astrof\1sica,
       Universidad Nacional Aut\'onoma de M\'exico, Campus Morelia \\
       Apartado Postal 3-72, 58090, Morelia, Michoac\'an, M\'exico}


\begin{abstract}

We report on the filaments that develop self-consistently in a new
numerical simulation of cloud formation by colliding flows.
As in previous studies, the forming cloud begins to
undergo gravitational collapse because it rapidly acquires a mass much
larger than the average Jeans mass.
Thus, the collapse soon becomes nearly pressureless, proceeding along
its shortest dimension first. This 
naturally produces filaments in the cloud, and clumps within the
filaments.
The filaments are not in equilibrium at any time, but instead
are long-lived flow features, through which the gas flows from the
cloud to the clumps.
The filaments are long-lived because they accrete {\it from}
their environment while simultaneously accreting {\it onto} the clumps
within them; they are essentially the locus where the flow changes from
accreting in two dimensions to accreting  in one dimension.
Moreover, the clumps also exhibit a hierarchical nature: the gas
in a filament flows onto a main, central clump, but other, smaller-scale
clumps form along the infalling gas. Correspondingly, the velocity along
the filament exhibits a hierarchy of jumps at the locations of the
clumps.
Two prominent
filaments in the simulation have lengths $\sim 15 \pc$, and masses $\sim
600\Msun$ above density $n \sim 10^3 \pcc$
($\sim 2 \times 10^3 \Msun$ at $n > 50 \pcc$).
The density profile exhibits a
central flattened core of size $\sim 0.3 \pc$ and an envelope that
decays as $r^{-2.5}$, in reasonable agreement with observations.
Accretion onto the filament reaches a maximum linear
density rate of $\sim 30 \Msun \Myr^{-1} \pc^{-1}$.

\end{abstract}
 

\section{Introduction}
\label{sec:intro}

Filamentary structure is ubiquitous in molecular clouds, both actively
star forming and quiescent \citep[e.g.,][] {Bally+87, Feitzinger+87,
Gutermuth+08, Myers2009, Juvela+09, Andre2010, Henning+10,
Menshchikov+10, Molinari+10, Schneider+10, Arzoumanian2011,
Kirk+13}. Moreover, these works have shown that, regardless of whether
the filaments are actively forming stars or not, they often contain
dense clumps and cores, which may or may not harbor young stellar
objects. If not, the cores are labeled ``prestellar''.

Numerous physical mechanisms have been advocated for the formation of
the filaments, such as the gravitational instability of a flattened
isothermal cloud, with and without magnetic fields \citep[e.g.,] []
{Larson85, Miyama+87, Gehman+96, Nagai+98, Curry00, Balsara+01};
supersonic MHD isothermal turbulence \citep[e.g.,][] {Padoan+01,
Goldsmith+08, Padoan+01}; and turbulence in thermally unstable gas
\citep[e.g.,][] {KW98, VS+06, WF00, Pavlovski+02}.
However, recent numerical simulations of cloud formation
and evolution including self-gravity systematically show that the
clouds, rather than remaining globally supported by turbulence, engage in
global gravitational contraction \citep[e.g.,] [] {BH04, HB07,
VazquezSemadeni2007, Vazquez-Semadeni2009, VS+11, HH08, Heitsch+08a,
Banerjee+09, Carroll+14}\footnote{\bf There have been some claims in the
  literature that the
  molecular clouds may be gravitationally unbound \citep[e.g.,] []
  {Dobbs+11}. However, closer examination of the evidence suggests that
  the clouds may become unbound as a consequence of the stellar feedback
  \citep[see the discussion by] [] {Colin+13}, so that the unbound stage is a {\it
  late} state of the clouds, while here we are interested in the {\it
  initial} conditions in the clouds.}. 
These results suggest that, although the inter\-stellar medium is highly
turbulent, 
the dynamics and structure of star-forming molecular clouds,
and the filaments within them, may actually be
dominated by gravity.  

That cold clouds may be in a state of global
gravitational contraction should not come in as a surprise, since the formation
mechanism of cold clouds may quickly endow them with masses much larger
than their Jeans mass. Indeed, 
cold clouds are expected to form, at least in solar-neighborhood-like
environments, by means of large-scale, coherent compressions in 
the warm neutral medium (WNM), driven, for example, by {\bf
large (kpc)}-scale instabilities
{\bf (Jeans, Toomre, Parker, and combinations thereof)}
\citep[e.g.,][]{Elm87, Kim+02}, or {\bf else} by the generic turbulence in this
medium, which is typically transonic \citep[e.g.,][]{HT03}. Such
compressions in the WNM are expected to induce a sudden phase transition
to the cold neutral medium\footnote{Note that this mechanism is not
in contradiction with the fact that observations \citep[e.g.,] []
{Dickey+77, Kalberla+85, SF95, HT03, JT11} as well as simulations
\citep[e.g.,] [] {VGS00, Gazol+01, Gazol+05, AB04, WN07, TB08}
systematically show that the density, temperature, and pressure in the
ISM span a continuum, rather than being restricted to narrow ranges of
values, as in the classical multi-phase models \citep{Field+69,
MO77}. As discussed by
\citet{VS+06}, the flow may {\it locally} undergo such phase transitions
whenever it is subject to compressions that push it out of thermal
equilibrium, creating condensation fronts that separate cold clouds from
their warm, diffuse environment. However, because there are significant
pressure fluctuations throughout the medium, the phase transitions will
occur around different mean pressures, and therefore the physical
conditions of the local warm and cold phases will differ from those of
the corresponding phases at a different location, producing a
statistical continuum of densities and temperatures
\citep{VS12b}. Moreover, \citet{Banerjee+09} showed that at some
locations the boundaries between the cold and warm phases are rather
sharp, while at other locations, they are significantly smoother,
depending on the local turbulent conditions, again producing significant
amounts of thermally unstable gas, which
  populate the density PDF in the classically forbidden regimes.}
\citep[CNM; e.g., ] [] {HP99, KI00, KI02,
Heitsch+05, VS+06}, which has a density roughly $100\times$ larger and a
temperature roughly $100\times$ lower than those of the WNM. Thus, the
Jeans mass, proportional to $\rho^{-1/2} T^{3/2}$, drops by a factor
$\sim 10^4$, and so the gas can easily acquire a large number of
Jeans masses upon the phase transition from the warm to the cold
phase. This implies that the ensuing collapse must occur in an
essentially pressureless manner.
However, if the collapse occurs in a nearly pressureless form, an
important consequence follows: as the collapse proceeds, it is expected
to amplify any deviations from isotropy, since it has been known for a
long time that pressureless triaxial spheroids collapse first along
their shortest dimension \citep{Lin+65}, producing first sheets and then
filaments. 

Interestingly, then, the formation of structures in cold,
dense clouds may be qualitatively similar to that in the so-called
``cosmic web'', which is believed to be permeating intergalactic space
\citep[e.g.,] [] {Bond+96, Cantalupo+14}, and which consists of a
network of filaments with dense galaxy clusters within them. In the
cosmic web, the filamentary structures are certainly not expected to be
in equilibrium, since they are actually undergoing gravitational
collapse.  Quite the opposite, the filaments feed material to the
clusters sitting within them and play a crucial role on the evolution of
the galaxies in the clusters.  Since observations of nearby molecular
clouds \citep{Myers2009} as well as
recent simulations of molecular cloud formation
\citep[e.g.,][]{VazquezSemadeni2007, VS+11, Heitsch+09, Naranjo+12} show
a similar hub-filament structure,
it is tempting to infer that the physical process in the cold
clouds is similar as well, implying that
their filamentary environment might have a strong influence on the
evolution of pre-stellar cores \citep[see also][]{Gomez2007, Smith+12}.


It is important to point out, however, that, contrary to the
cosmological case, a nascent cold cloud is internally turbulent, due to
several fluid instabilities acting during its formation process
\citep{Hunter+86, Vishniac94, KI02, AH05, Heitsch+05, VS+06}. It has
been thought for some time that this turbulence may be strong enough to
support the clouds \citep[e.g.] [] {KH10}, and to simultaneously induce
local compressions in which the local Jeans mass is reduced sufficently
to cause the local fluctuation to undergo collapse \citep{Padoan95,
PN02, PN11, VS+03b, KM05, HC11}. However, early on it was pointed out by
\citet{CB05} that generally, the turbulent compressions alone do not
provide a sufficent reduction of the local Jeans mass as to directly
induce local collapse. This is supported by the fact that numerical
simulations of cloud formation indicate that the turbulence generated as
the cloud is assembled is only moderately, rather than strongly,
supersonic \citep[e.g.] [] {KI02, AH05, Heitsch+05,
VS+06,Vazquez-Semadeni2010, 
Banerjee+09}, and that {\it local} gravitational collapse does not start
until several million years later in the clouds' evolution, after
significant contraction of the clouds at large \citep[e.g.,]
[] {VazquezSemadeni2007, Vazquez-Semadeni2010, Heitsch+08a}. 


However, once the nonlinear density fluctuations begin to collapse
locally, they do so on shorter timescales than that of the whole cloud,
precisely because of their nonlinearly larger local density compared to
the average cloud density, causing a regime of hierarchical
gravitational fragmentation \citep[i.e., of collapses within collapses;
][]{Vazquez-Semadeni2009}, similar to the regime originally proposal of
\citet{Hoyle53}, except for the nonlinearity of the density
fluctuations \citep[see, e.g., the review by][]{VS12}. 

Note also that, as originally pointed out by
\citet{Hoyle53}, if the collapse is nearly isothermal, then number of
Jeans masses in a cloud increases as it collapses. This implies that
the collapse of the cloud continues to behave as a pressure-free
flow. Although the thermal pressure does indeed increase during
the collapse, it is always lagging behind the gravitational energy
density, and by an ever-larger margin. In this way, the collapse of a
turbulent, isothermal cloud is seen to behave similarly to the
collisionless dark matter fluid
and thus it is expected to proceed by enhancing its eccentricity,
effectively collapsing first along its shortest dimension, and thus
forming first sheets and then filaments \citep{Lin+65}. The filaments
formed by this global process are expected to be surrounded by this
accretion-driven {\it ram}-pressure, and dynamically very different from
filaments confined by the thermal pressure of a static medium.

Recent observational studies have emphasized the filamentary 
structure of molecular clouds, mainly thanks to the advent of 
the {\it Herschel} telescope, devoting special attention to the
spatial and velocity structure of the filaments, and the implied
accretion rates onto and along the filaments
\citep{Andre2010,Arzoumanian2011,Battersby+14,Hacar2013,
Kirk+13,Molinari+10,Myers2009}.
With this in mind,
in this paper, we present a new, high-resolution numerical simulation of
dense cloud formation by colliding flows, and report the physical
conditions of the filaments that develop in this simulation by
gravitational contraction. We have chosen this setup as
it provides a self-consistent mechanism for driving the turbulence in
the forming dense cloud, contrary to the case of simulations in which
the turbulence {\bf is} externally driven by a force generated in Fourier space
and applied everywhere in the flow. The latter scheme, although quite
common in the literature \citep[e.g.,] [] {PN99, Padoan+07, KHM00,
VS+03b, VS+05, VS+08, Kritsuk+07, FK12, FK13}, has the disadvantage that the
turbulent Mach number is imposed, rather than self-consistently
produced, and thus it is not possible to know whether it is 
realistic or not. An exceesive imposed turbulent Mach number can indeed have
the effect of providing global support while simultaneously inducing
local collapse, eliminating the feature of global collapse, but this
is likely to be an artifact of the excessive and whole-volume-application
nature of the driving.

The plan of the paper is as follows: in
\S\ref{sec:model} we present the numerical simulation, to then
discuss its global evolution in \S\ref{sec:global_evol}. Next, in
\S\ref{sec:filaments} we report the physical properties and
conditions in one fiducial filament that develops in the simulation.
Finally, in \S\ref{sec:summary}, we present a summary and our
conclusions.

\section{Numerical model}
\label{sec:model}

The simulation used in this study is similar to the simulation
L256$\Delta v$0.17 discussed in \citet[][hereafter Paper I]
{VazquezSemadeni2007}, although performed at a significantly higher
resolution ($296^3 \approx 2.6 \times 10^7$ SPH particles {\it versus}
$148^3 \approx 3.2 \times 10^6$), and performed with the {\sc gadget-2}
code, while in Paper I the code {\sc gadget} was used. The
simulation describes the collision of two oppositely-directed streams of
warm gas. The collision triggers a transition to the cold phase, forming
a cloud that rapidly grows in mass by accretion of the warm, diffuse
inflow material until it becomes Jeans unstable and begins to
collapse. We refer the reader to Paper I for specific details of the
setup and a description of the more general behavior. While the
general behavior of the simulation is the same as in Paper I, the higher
resolution causes some differences in the details of the evolution,
which we describe below.


The simulation was performed in a box of $256\pc$ per side, initially
filled with warm gas of uniform density ($1~\pcc$) and temperature
($T=5206$ K). The box thus contains a total mass of $5.26 \times 10^7
\Msun$.  At our chosen number of SPH particles, assuming a mass per
gas particle of $m_\eff = 2.12\times 10^{-24}\gr$, the mass per SPH
particle in our simulation is $\approx 0.02\Msun$.

Two cylindrical regions within this box were set to collide within this
box, with initial velocities $v_{{\rm inf}}$ ($=9.2 \kms= 1.22 c_s$,
where $c_s$ is the adiabatic sound speed), and $-v_{{\rm inf}}$, so that
they collide at the central $(y,z)$ plane of the box.  Note that
the inflows have the same density as the ambient medium, so they are
only distinguished from it by their velocity, not by their density.

The dimensions of the cylinders
were $l_{{\rm inf}}=112\pc$ and $r_{{\rm inf}}=32\pc$,
and thus contained $2.64 \times 10^4 \Msun$
each, causing the total mass in the cold cloud to be $\gtrsim 5.2 \times
10^4 \Msun$. Note that the mass is larger than twice the mass in the inflows
because the latter entrain some ambient material that is ultimately
accreted onto the cloud.  

In addition to these flows, small-amplitude
turbulent motions were added to the gas, in order to break the symmetry
of the setup and allow the development of instabilities in the dense
layer formed by the collision (the cloud).  As in Paper I, the code was
modified to model the formation of sink particles. Specifically, in
order to prevent the timestep from becoming arbitrarily small when a
certain region of the flow undergoes gravitational collapse, once a
region becomes denser than a certain threshold $\nsf$, it undergoes a
number of tests to determine whether it is collapsing. If the tests are
passed, then the mass in the region is removed from the gas phase and a
``sink particle'' is created, which inherits the total mass and momentum
of the removed SPH particles \citep{Bate+95, Jappsen+05,
Federrath+10}. In turn, the sink particle itself may continue to accrete
mass from the gas phase. In the present simulation, we have taken $\nsf
= 3.2 \times 10^6
\pcc$, the same as that in Paper I.\footnote{Note that in Paper I
there was a typographical error, with the threshold density erroneously
reported as $10^5 \pcc$.} In addition, we also implemented the
radiative heating and cooling function used in Paper I, which in turn
was taken from
\citet{KI02}, only correcting some typographical errors.
The effect of more realistic cooling functions has been
recently investigated by \citet{Micic+13}, showing that they are in
general minor.

The collision of the flows in the center of the box induces a phase
transition in the flow from diffuse-warm to cold-dense gas, producing a
dense layer of cold gas that grows in mass over time \citep{VS+06}. The
ram pressure-confined dense layer is subject to bending-mode and
hydrodynamical instabilities, causing the development of moderately
supersonic turbulence within it \citep{Vishniac94, WF00, Heitsch+05,
Heitsch+06, VS+06}, which in turn produces nonlinear density
fluctuations.  Moreover, the dense layer continues to accrete gas from
the original inflows, thus increasing its mass.
Eventually, as was shown in Paper I, the layer at large becomes
gravitationally unstable and begins to contract. Some
time later, before the global contraction is completed, the local
density fluctuations also become unstable due to the decrease of
the average Jeans mass, and begin to collapse themselves. Because they are
nonlinearly overdense with respect to the global cloud average, they
have shorter free-fall times than that of the cloud at large, completing
their collapse before the whole cloud does. 

\begin{figure*}[!t]
  \includegraphics[width=0.48\textwidth]{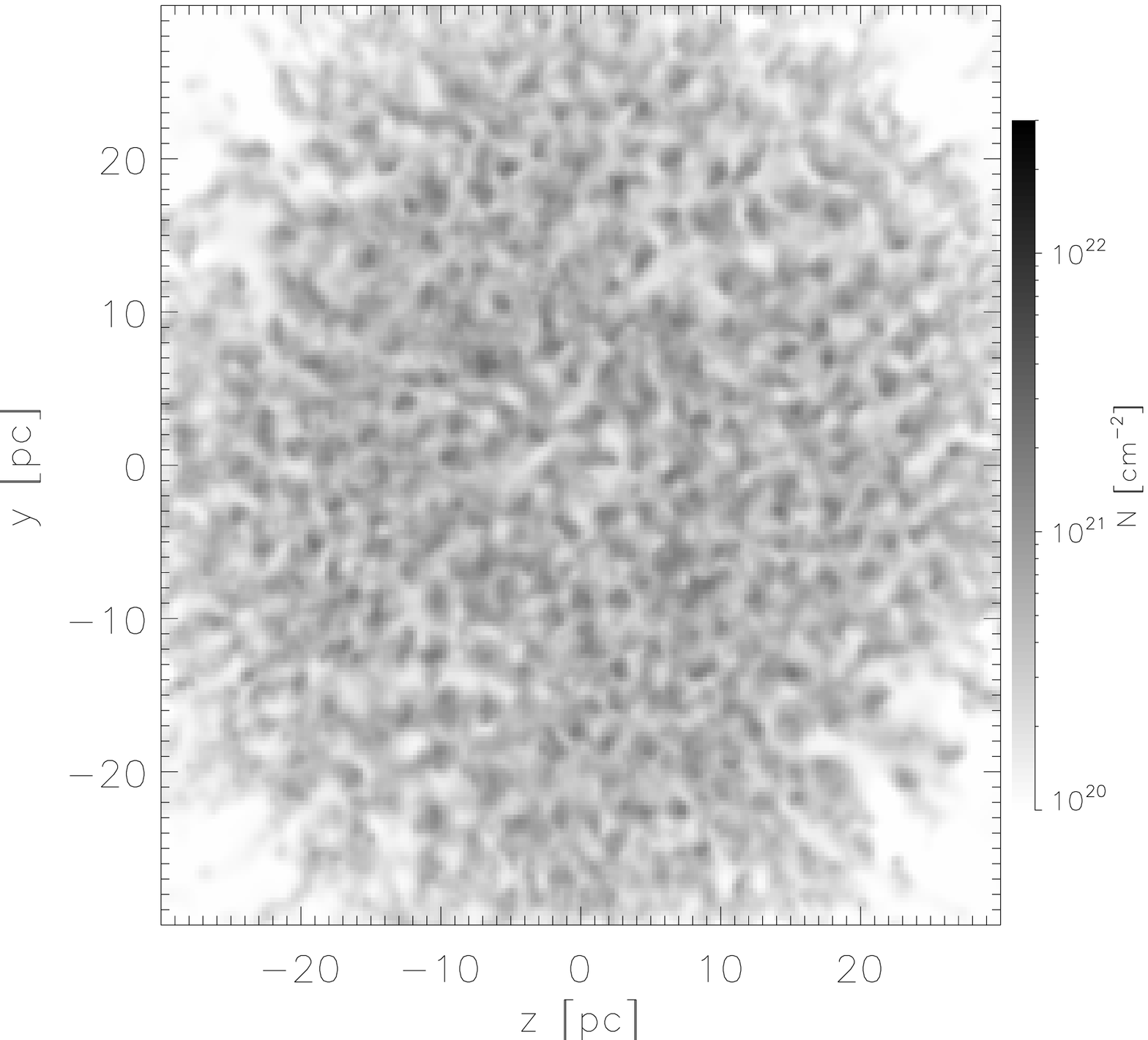}
  \includegraphics[width=0.48\textwidth]{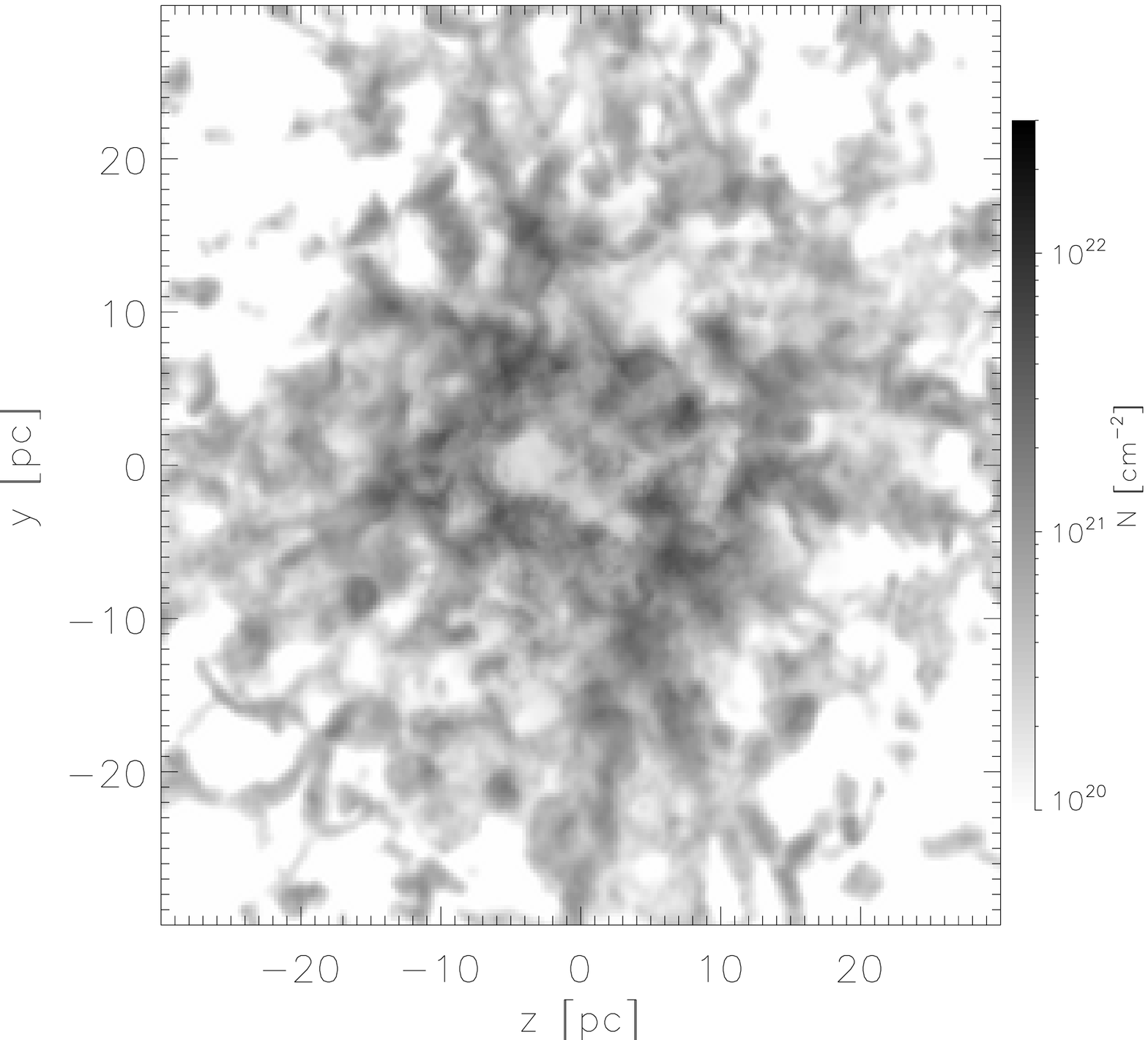}

  \includegraphics[width=0.48\textwidth]{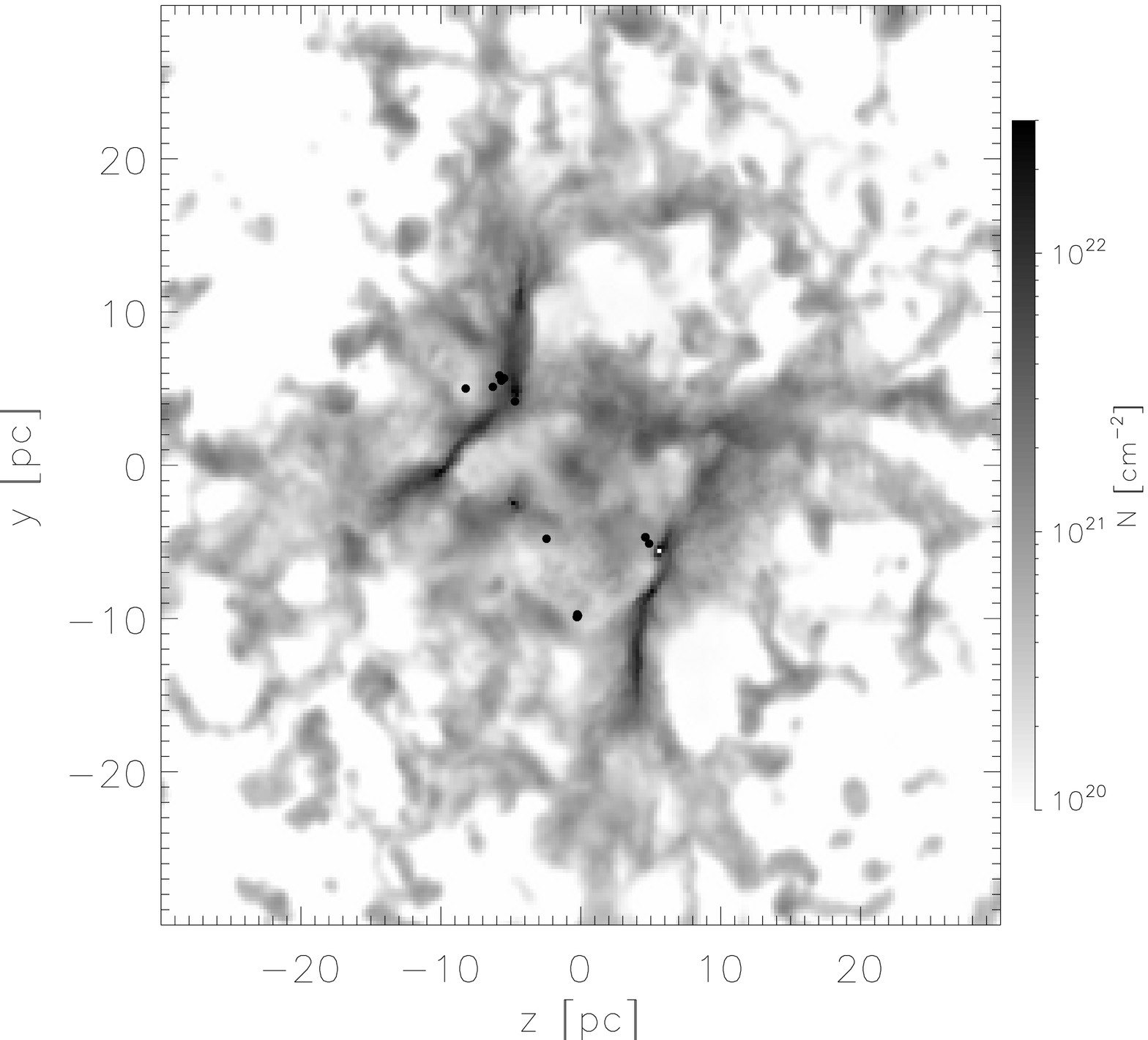}
  \includegraphics[width=0.48\textwidth]{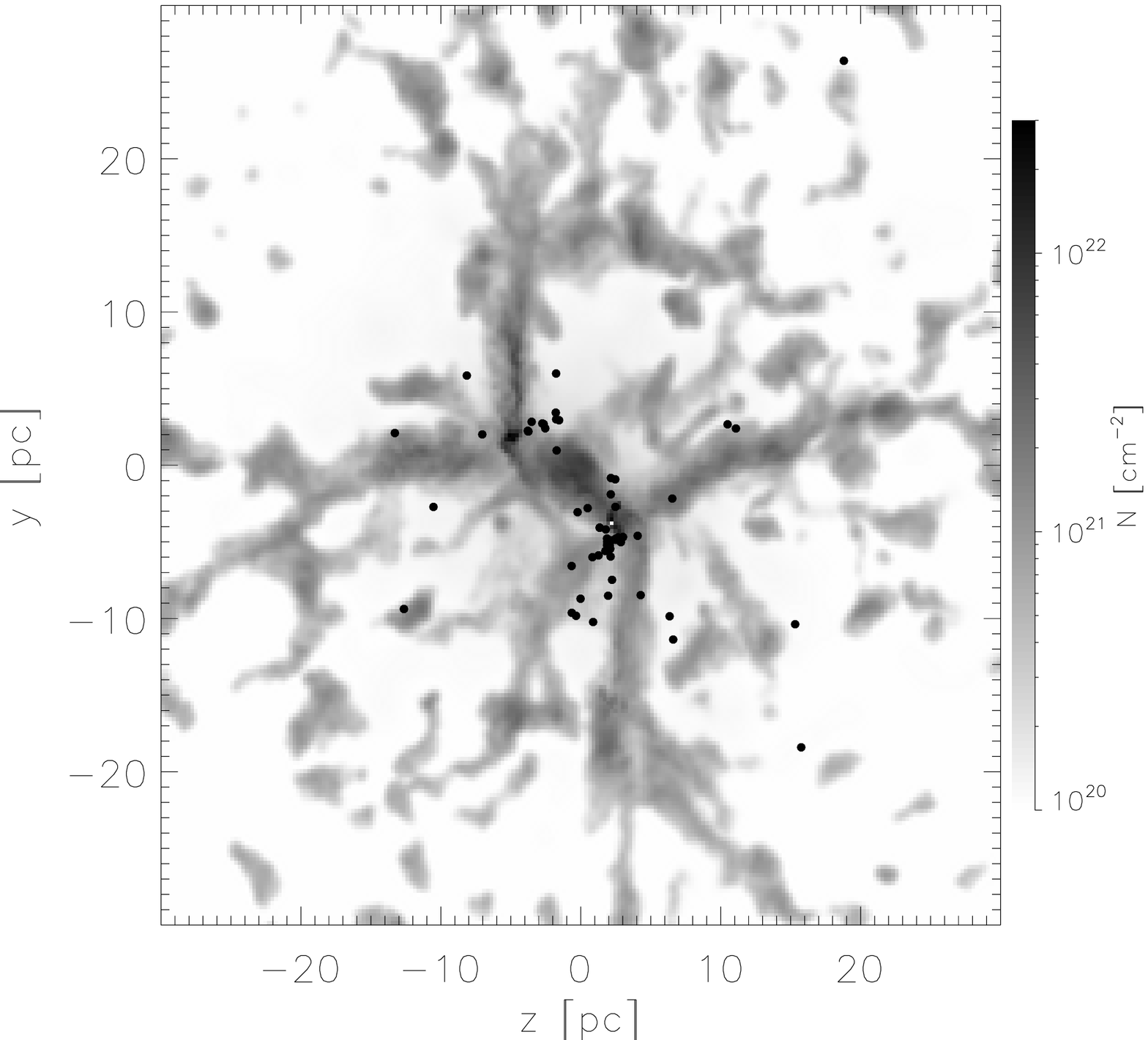}
  \caption{Face-on view of the central $60\pc$ of the simulation
    along the axis of the colliding flows.
    Grayscale represents the gas column density integrated over
    $\pm 15\pc$ around the contact point of the initial flows, at
    the center of the simulation box.
    Times shown correspond to
    $t=9.30$ ({\it top-left panel}),
    $17.93$ ({\it top-right}),
    $23.91$ ({\it bottom-left}) and
    $30.41 \Myr$ ({\it bottom-right})
    from the onset of the simulation.
    Dots indicate the positions of the sink particles in the range
    depicted.
    Coordinates are measured from the center of the simulation box.}
  \label{fig:faceon}
\end{figure*}


\section{Global evolution} \label{sec:global_evol}

Paper I aimed at investigating the combined effects of cooling,
turbulence generation in the compressed layer (the cloud), and
self-gravity, finding that the cloud grows in mass until its
gravitational energy overtakes the turbulent kinetic energy, at which
time it begins to undergo gravitational collapse. However, the turbulent
kinetic energy induced by the instabilities in the compressed layer was
not very large \citep[the Mach number was of the order of a few only;
see also][]{Vazquez-Semadeni2010}, and velocities corresponding to Mach
numbers typical of giant molecular clouds only developed as a
consequence of the gravitational collapse, and corresponded to the
collapsing motions themselves, not to random turbulence. Several Myr
after the onset of global collapse, local collapses began to occur,
seeded by the density fluctuations produced by the moderately supersonic
turbulence produced by the flow collision. 

However, in the main simulation of Paper I, of which the present
simulation is a remake at higher resolution, most of the
early star formation occurred in a high density ring that formed at the
circular periphery of the flat cloud caused by the inflowing cylinders.
Instead, in the higher-resolution run presented here, the lower
numerical viscosity allows the gas to squirt out more freely from the
colliding region, so that the high density ring does not form (see fig.\
\ref{fig:faceon}). This has two implications.
First, while in the run from Paper I all the
initial local collapses occurred in the peripheral ring, in the present
simulation they occur throughout the dense layer. Second, while in Paper
I the ring itself quickly began contracting and falling onto the center
of the dense layer, here, in the absence of such a ring, both the local
and global collapses are less focused, and thus take longer to develop.
So, while sink formation started at $t \approx 17\Myr$ in Paper I, here
it begins at $t \approx 20\Myr$.

Also, the global collapse is less coherent. While in Paper I the time at
which the ring reaches the simulation center is very well defined,
signaling the culmination of the large-scale collapse, here the dense
layer fragments into filaments, which then form clumps, onto which the
rest of the filament continues to accrete. This is understood in terms
of the fact that the collapse timescale for a long filament of aspect
ratio $A$ is longer
than that of a roughly spherical clump of the same volume density by a
factor between $\sim A^{1/2}$ and $\sim A$, depending on whether the
collapse is homologous or proceeds outside-in
\citep{Toala, Pon+12}. Only later, the entire filament-clump system
falls into the large-scale potential well at the center of the
simulation. Specifically, the free-fall time at the mean
initial density of cloud after the phase transition ($\sim 100 \pcc$) is
$\sim 3.3$ Myr. However, in practice, the global collapse of the cloud occurs
over roughly 20 Myr (see Paper I and Fig.\  \ref{fig:faceon}), from $t
\sim 13$ Myr to $t \sim 33$ Myr.

The fact that the usage of higher resolution (and perhaps also the
different version of the code) produce a different morphological pattern
is a reflection of the chaotic nature of the system, which exhibits
sensitivity to initial conditions (i.e., arbitrarily nearby initial
conditions end up producing very different states after finite evolution
times). However, the differences occur only at the level of morphological
detail, and with timing differences at the $\sim 15$\% level.
Statistically, the two simulations behave in a similar manner,
developing local collapses that culminate earlier than the global
collapse.

It is important to remark that the collision of WNM currents, aided by
thermal instability, initially forms a planar structure, or sheet-like
cloud, which rapidly becomes Jeans-unstable and breaks into filamentary
structures. In parallel, the sheet keeps accreting material from the
inflows, implying that the amount of gas that has undergone the phase
transition into CNM (the cloud's mass) increases in time.  In a similar
way, once formed, the filaments accrete mass from the larger planar
structure, increasing their mass until they become gravitationally
unstable; the filaments then collapse as a whole. But, in turn, the
filaments fragment into a number of clumps because they have a longer
free-fall time than any roughly spherical, Jeans-unstable fluctuation
within them \citep{Toala, Pon+12}. Thus, the clumps within the filaments
increase their own density and develop a density contrast with respect
to their parent filament. Material from the filament continues to ``rain
down'' (accrete) onto the clumps, while continuing to be fed by the
accretion flow from the sheetlike cloud. 

The whole system thus constitutes a mass cascade from the largest to the
smallest scales in the cloud, similar to that envisioned by
\citet{Field+08}, except that those authors proposed a process in which
the energy released by the collapse at each scale produces random,
quasi-isotropic motions at a smaller scale (see their Sec.\ 3). Instead,
what we observe in our simulation is a continuous gravitational collapse
all the way down to the protostar scale, as observed, for example, by
\citet{Galvan+09}, \citet{Schneider+10}, and \citet{Csengeri+11}. Also,
a distinctive feature of this cascade is that the structures at
different scales have different morphologies: from sheets to filaments,
and from filaments to clumps.

A crucial feature of these structures is that {\it none of them are in
equilibrium.} Instead, they represent a continuous flow towards the
troughs of the gravitational potential, at both local and global scales,
giving rise to a regime of collapses within collapses, or {\it
hierarchical gravitational fragmentation}
\citep{Vazquez-Semadeni2009}. Structures are actually {\it 
features of the flow} that persist because they continuously being fed
by the accretion flow.
One very important kind of such features are the
filaments, which recently have received much observational attention. 
In Fig.\ \ref{fig:faceon} we show four snapshots of the simulation
viewed face-on, showing the formation of the filaments and then their
merging to form a final dense collapse center. In
the remainder of the paper, we now discuss the physical properties of
the filaments that develop in our simulation.

It is important to point out that it is difficult to
assign a well-defined timescale to the filaments, because typically they
last longer than their crossing time, due to the accretion.  Moreover,
they take time to assemble and to take a well defined shape. In any case,
the filaments we discuss in the next section roughly last $\sim 7$ Myr,
from $t \sim 21$ to $t \sim 28$ Myr, from the time when they first become
discernible to the time when they finish accreting onto their nearest
main clump. We choose to study them at a specific time when they appear
clearly defined.

\section{Dense filaments}
\label{sec:filaments}

\subsection{Identification} \label{sec:ident}

\begin{figure}
  \includegraphics[width=0.48\textwidth]{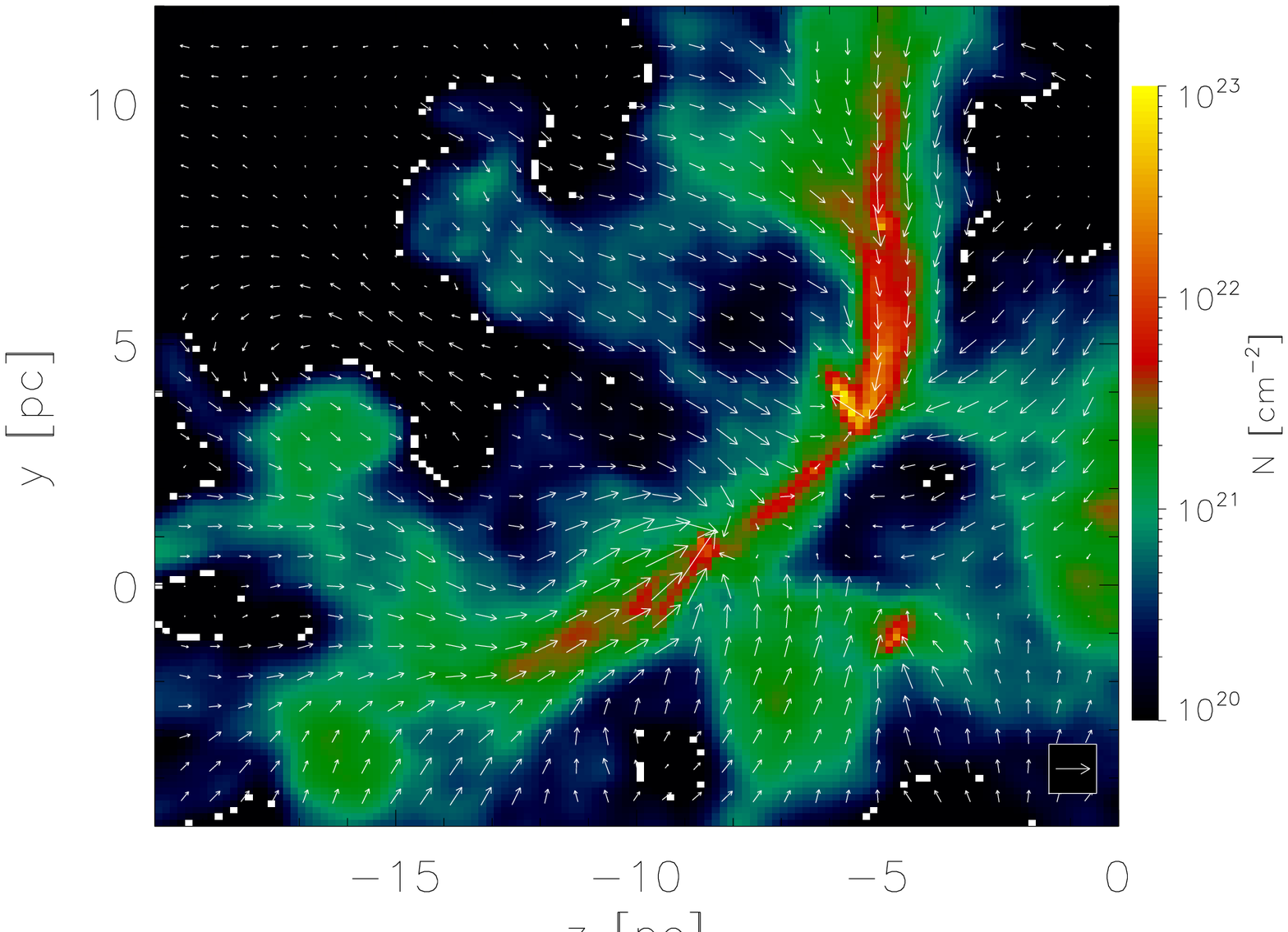}

  \includegraphics[width=0.48\textwidth]{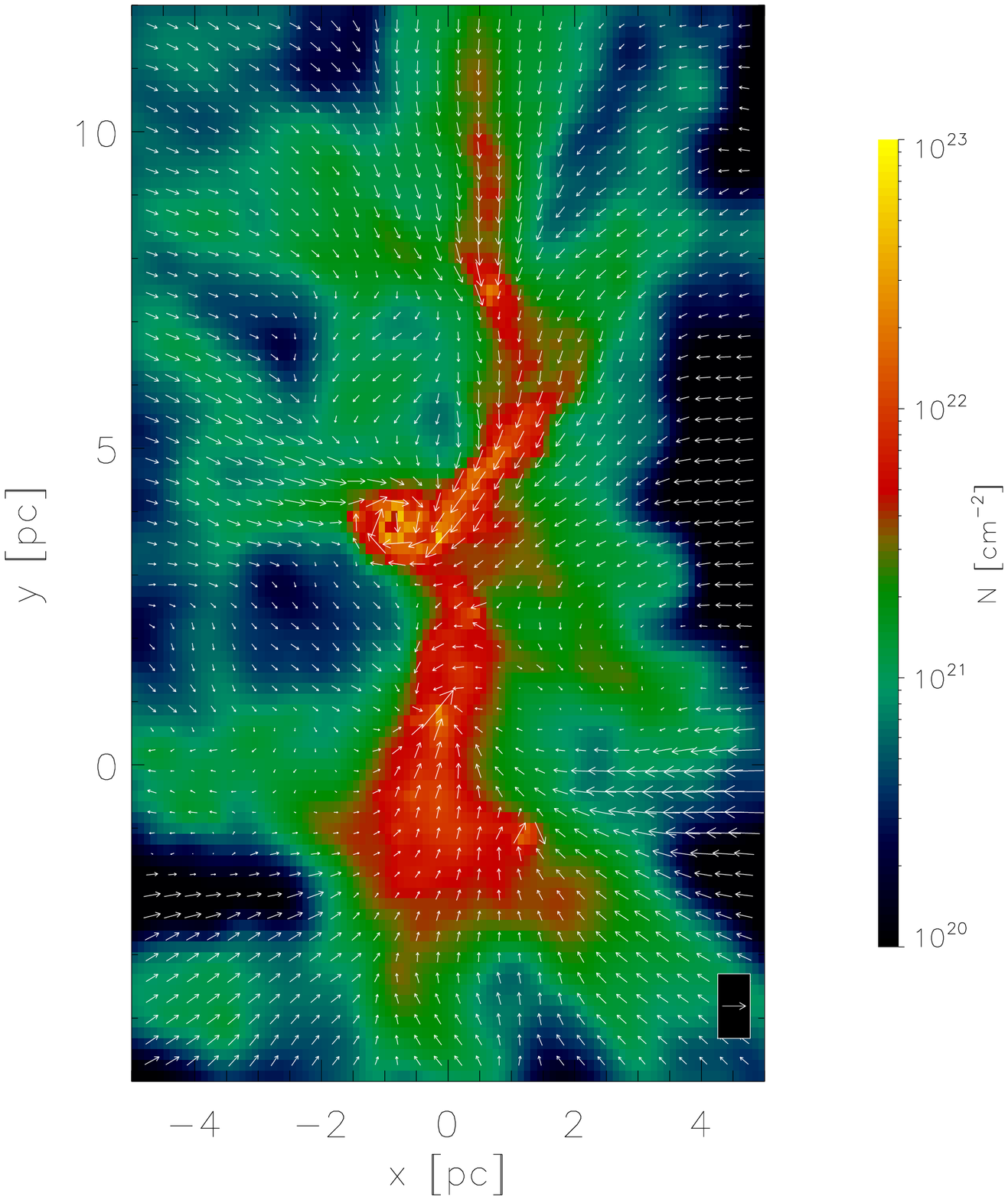}
  \caption{Filament 1 at $26.56\Myr$ into the simu\-la\-tion.
    Colors show the column density of the gas integrated over the
    range $|x|<5\pc$ for the top panel, and $-20\pc<z<0$ for the
    bottom panel.
    Coordinates are measured with respect to the center of the
    simulation box.
    The arrows show the density-weighted projected velocity field,
    with the arrow in the lower right representing $2\kms$.
    See the electronic edition of the Journal for a color version
    of this figure.}
  \label{fig:fil1}
\end{figure}

  \begin{figure*}[!t]
    \hspace{0.6cm}
    \includegraphics[width=0.28\textwidth]{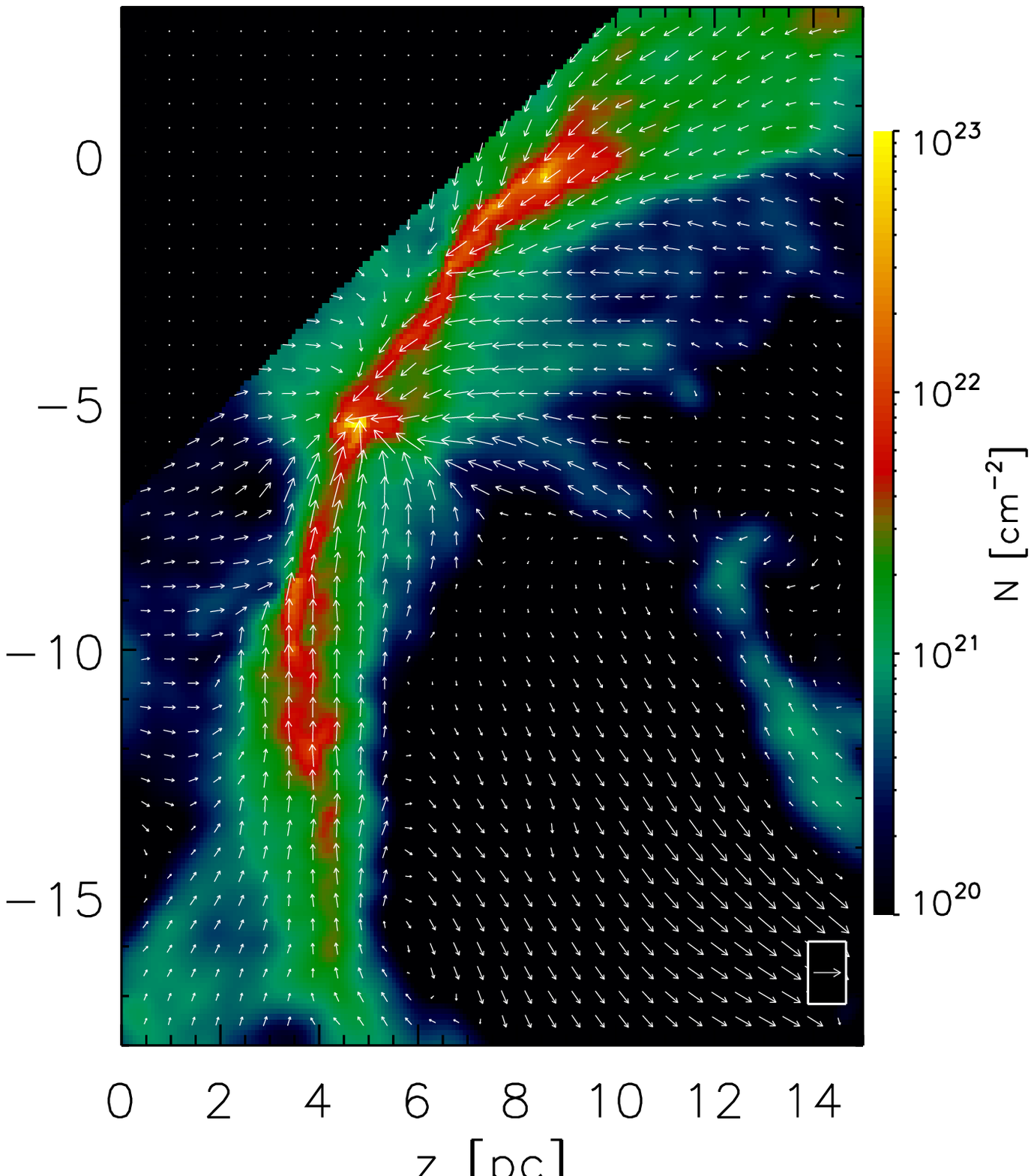}
    \hspace{0.6cm}
    \includegraphics[width=0.28\textwidth]{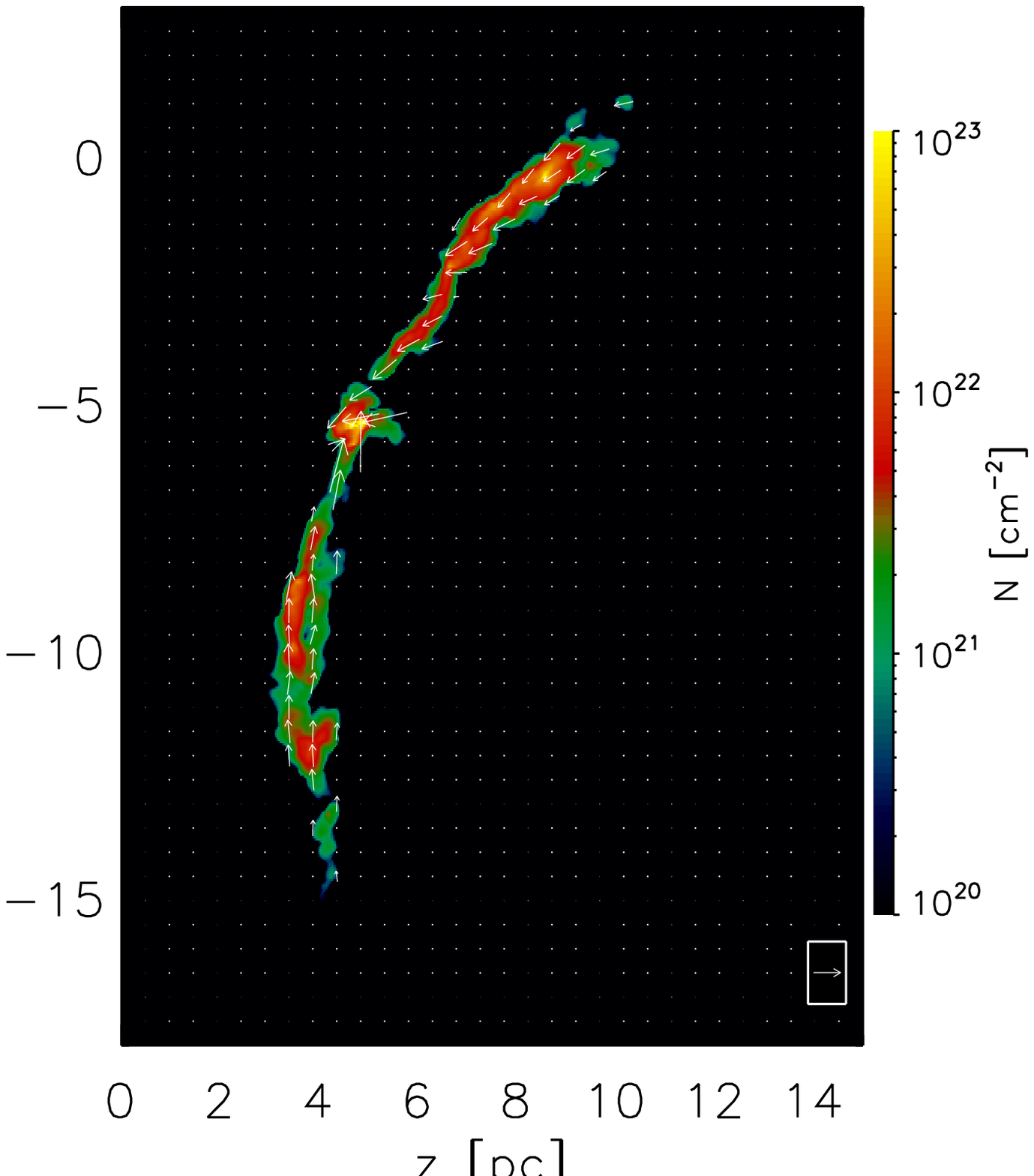}
    \hspace{0.6cm}
    \includegraphics[width=0.28\textwidth]{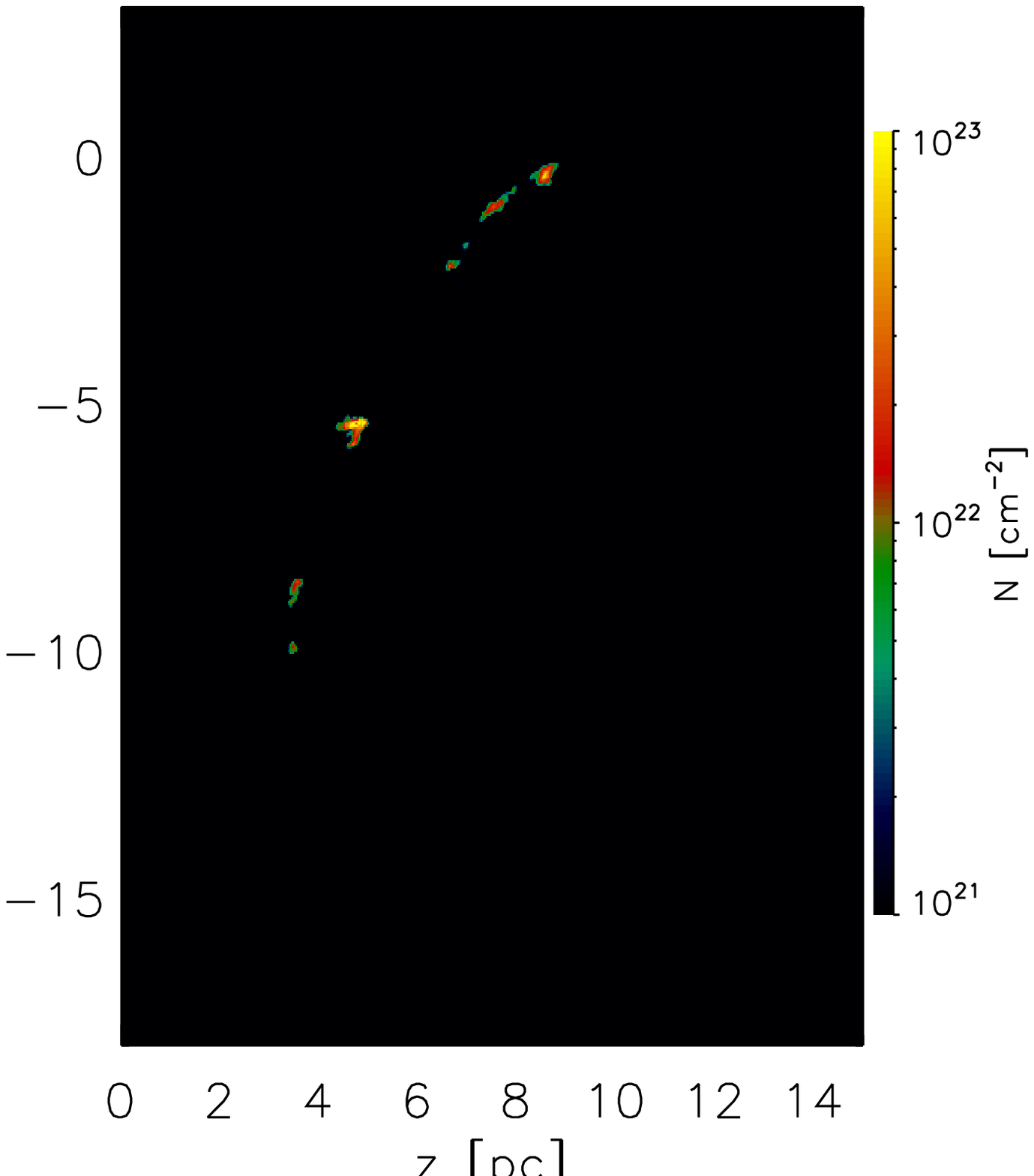}

    \hspace{0.6cm}
    \includegraphics[width=0.28\textwidth]{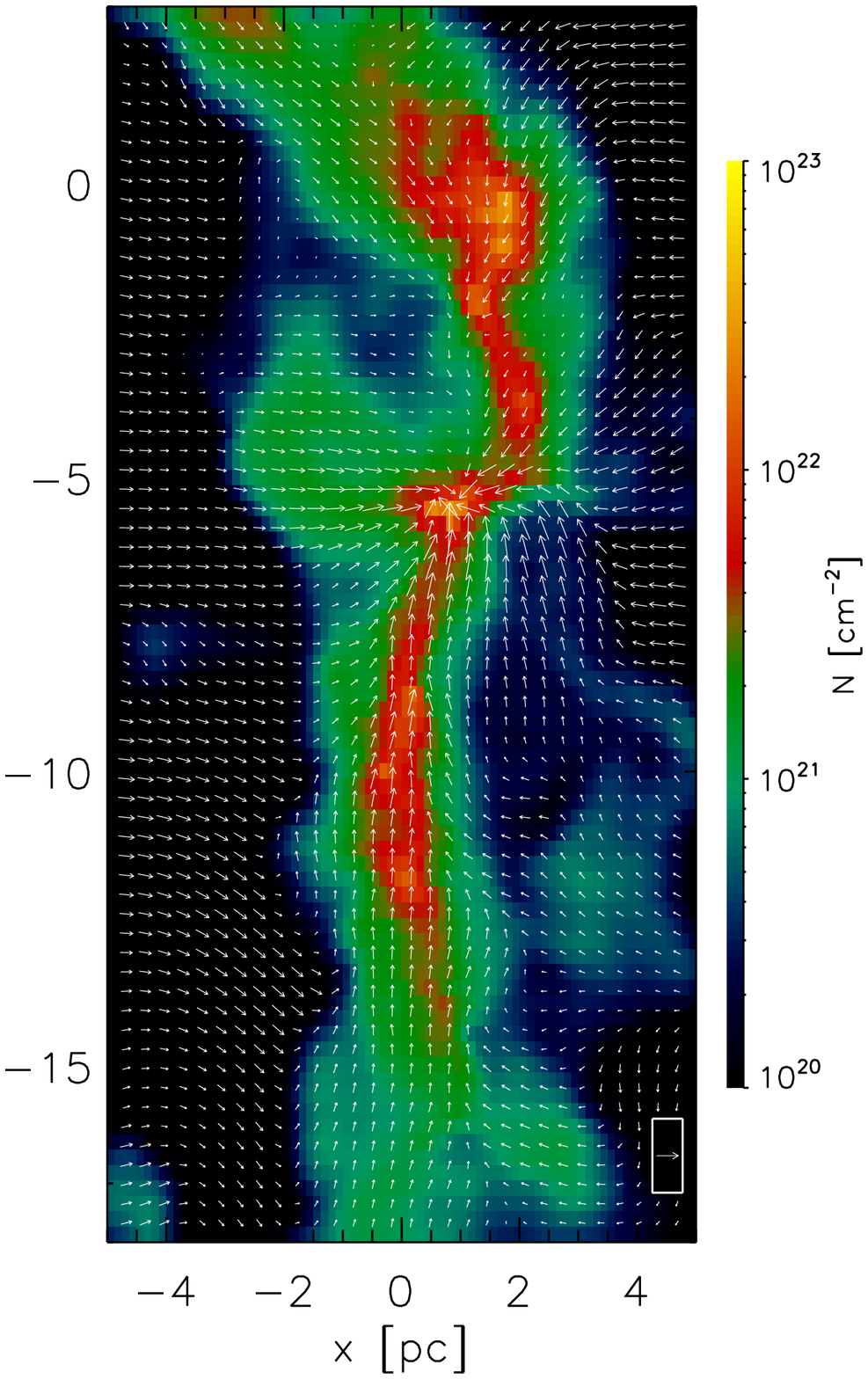}
    \hspace{0.6cm}
    \includegraphics[width=0.28\textwidth]{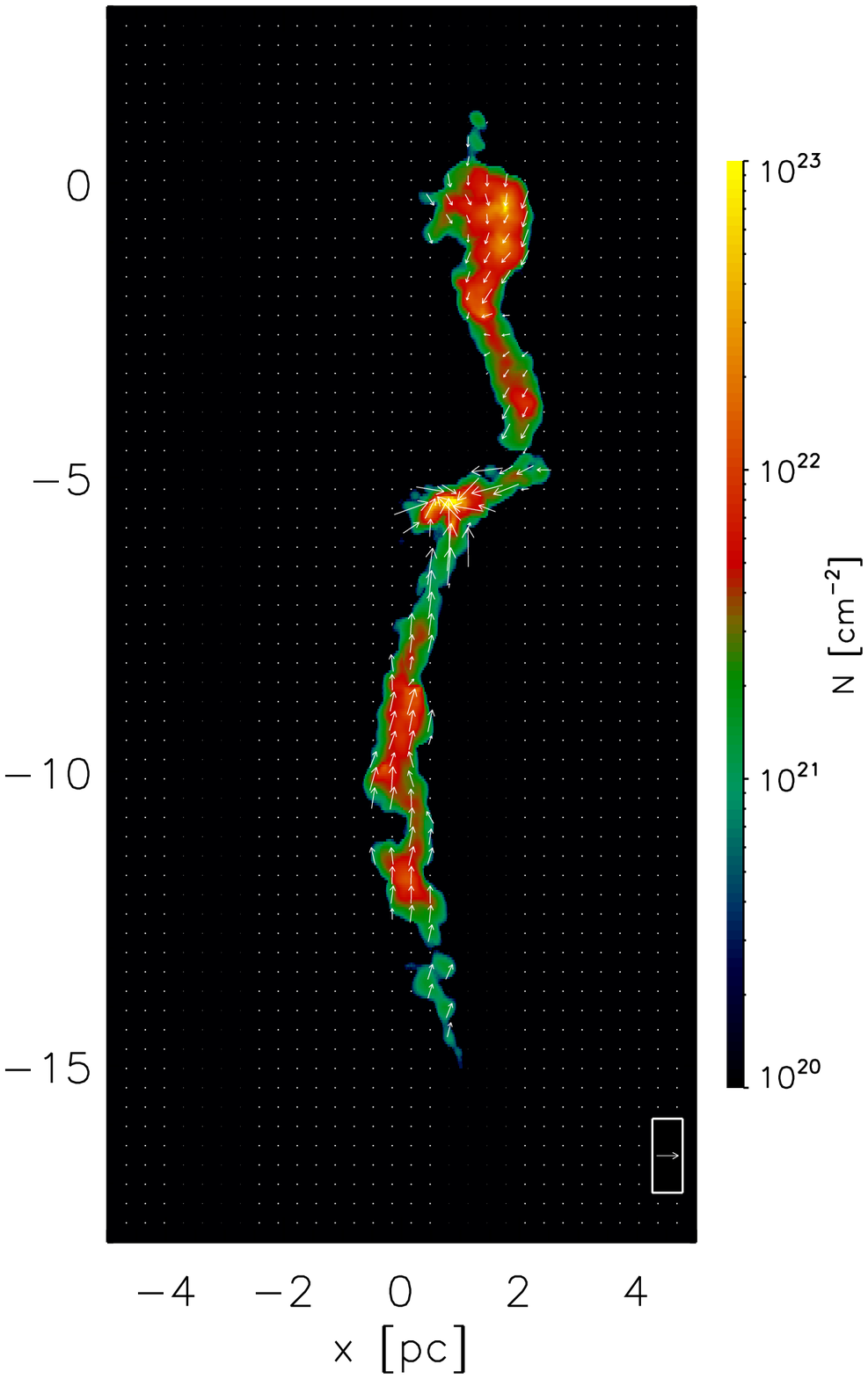}
    \hspace{0.6cm}
    \includegraphics[width=0.28\textwidth]{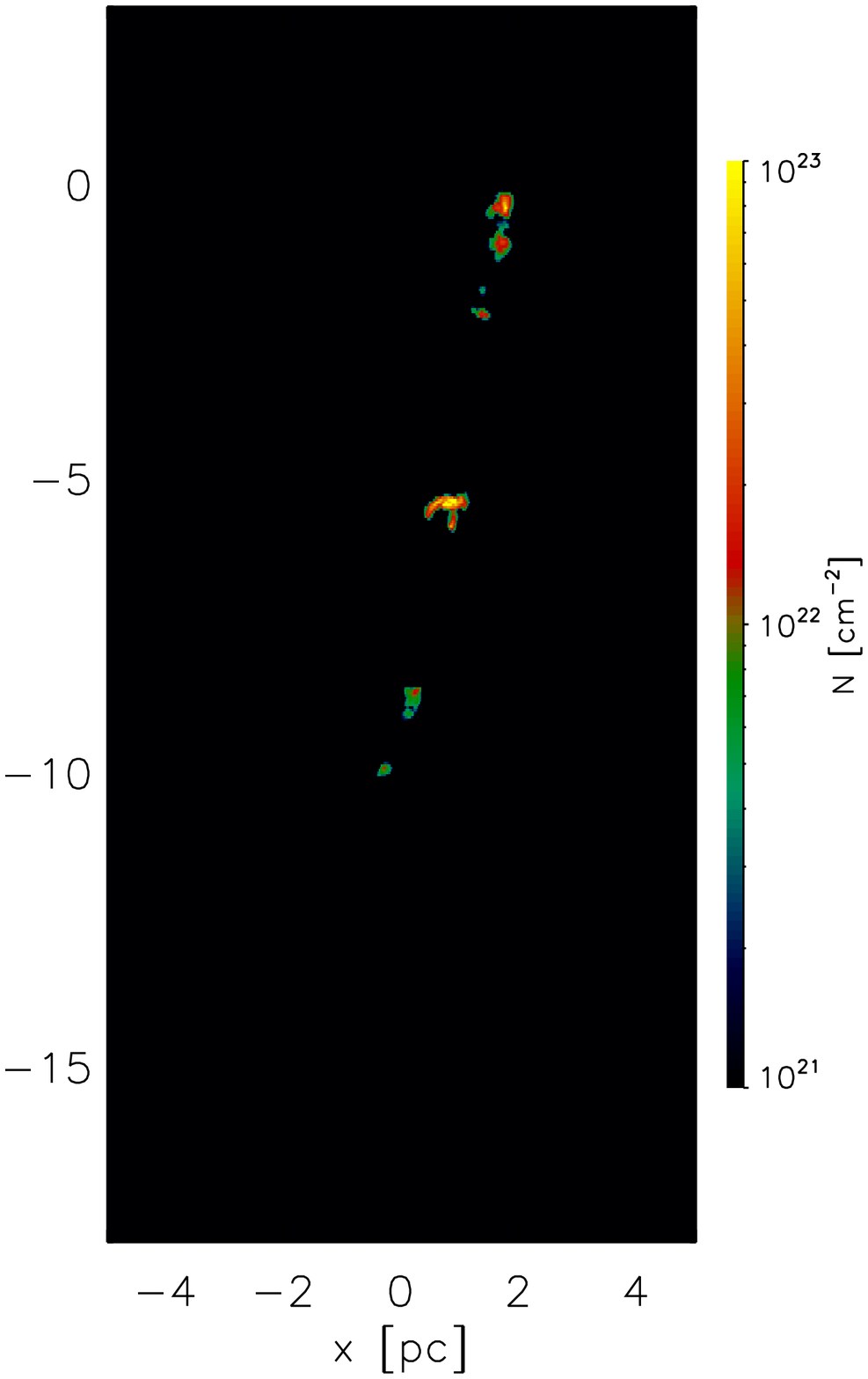}
    \caption{Filament 2 at $26.56\Myr$
      into the simulation, integrated in the ranges $|x|<5\pc$ for the top
      row, and $0<z<15\pc$ for the bottom row. The
      left column shows the total column density, while the middle
      column shows column density for gas with $n>10^3\pcc$, and
      $n>10^4\pcc$ in the right column.
      In all cases, the arrows show the density-weighted projected
      velocity, with the arrow in the lower right representing $2\kms$.
      The region $z-y \ge 7\pc$ is suppressed to avoid a separate
      condensation unrelated to the filament of interest.
      See the electronic edition of the Journal for a color version
      of this figure.
      The bottom row of this figure is also available as mpeg
      animations in the
      electronic edition of the {\it Astrophysical Journal}.}
    \label{fig:fil2}
  \end{figure*}

Two major filaments are readily noticeable in the lower left panel of
Fig.\ \ref{fig:faceon} ($t=17.9$ Myr);
we shall refer to the one centered at
$(z,y)\approx(-7.6,2.8)$ pc as ``Filament 1'', and to the one centered
at $(z,y)\approx(6.8,-5.3)\pc$ as  ``Filament 2''.
Figures \ref{fig:fil1} and \ref{fig:fil2} show column density images of
Filaments 1 and 2, 
respectively, in both the $(z,y)$ and the $(x,y)$ planes,
$26.56 \Myr$ into the simulation ({\bf approx}. the midpoint of the
times shown in the bottom panels of fig. \ref{fig:faceon}). The
coordinates shown in the axes are measured with respect to the
center of the simulation box.

When only gas with density $n > 10^3\pcc$ is considered,
Filament 1 has a mass $\sim 560\Msun$
($\sim 2.4\times10^3\Msun$ if we consider gas with $n > 50\pcc$),
and is $\sim 12\pc$ long and $\sim 1\pc$ wide.
Its mean density ($5.18 \times 10^4 \pcc$, mass-weighted
considering gas with $n > 10^3\pcc$ only) and aspect ratio imply an
approximate free-fall time of $0.72\Myr$ \citep{Toala}.
Filament 2 has a mass $\sim 680\Msun$ when defined by a threshold $n>10^3\pcc$
($\sim 2\times10^3\Msun$ when gas with $n>50\pcc$ is considered),
has an approximate length of $\sim 15\pc$, and a width $\sim 1\pc$;
its mean density ($7.88\times10^4\pcc$) and aspect ratio imply an
approximate free-fall time of $0.66\Myr$.

\subsection{Local gravitational collapse}
\label{sec:collapse}

As discussed in \S\ref{sec:global_evol}, the filaments soon develop
dense cores. By the time shown in Figs.\ \ref{fig:fil1} and
\ref{fig:fil2}, some of those collapse centers have already created sink
particles (35, totaling $2.24 \times 10^3 \Msun$). Thus, each of
the individual structures formed from the original collision of WNM
currents acts as a channel that funnels mass from the parent to the
child structures. Moreover, as time goes on, the scale of the
filament-clump system grows, involving increasing amounts of mass and
correspondingly larger spatial scales. For example, in the bottom-right
panel of Fig.\ \ref{fig:faceon} a large accreting clump is seen to have
formed at the center of the frame, with large-scale filaments accreting
onto it.

\begin{figure}[!t]
  \centerline{ \includegraphics[width=0.50\textwidth]{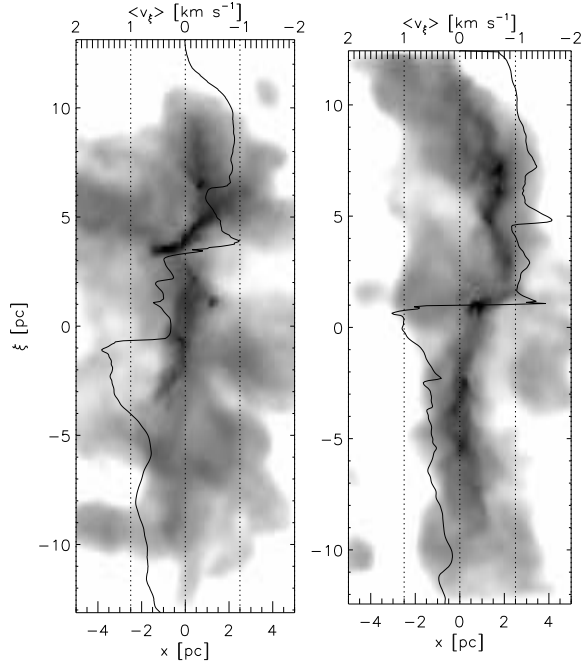} }
  \caption{Projections perpendicular to the filaments.  The grayscale
  shows the column density
  (in the $10^{20}$ to $3\times 10^{22} \cm^{-2}$ range,
  logarithmic scale)
  along a line of slope $-50^\circ$ in the
  $z$-$y$ plane ({\it top panel}) of Fig.\ \ref{fig:fil1} for Filament 1 ({\it
  left}), and $-20^\circ$ in the $z$-$y$ plane ({\it top panel}) of Fig.\
  \ref{fig:fil2} for Filament 2 ({\it right}), so that the vertical
  coordinate here (labeled $\xi$ and measured from the center of mass of
  each filament) is approximately along the filaments.  The solid line
  shows the column density-weighted $\xi$-velocity, averaged over the
  filament width. Its value is given by the upper horizontal scale.  The
  global collapse of the filaments, with superimposed local
  collapses, is apparent.  } \label{fig:vel-xi}
\end{figure}

The picture described above may be verified by exploring the velocity
field along the filaments. In order to do this, we have calculated the
column density along lines of sight (LOSs) approximately perpendicular
to the filaments.
We first obtained the density distribution of the gas on two
boxes with $512^3$ grid points,
each containing one of the filaments, with resulting
resolution of $(\Delta x,\Delta y,\Delta z) = (0.020,0.033,0.039)\pc$
for Filament 1, and $(0.020,0.041,0.029)\pc$ for Filament 2. Then,
taking advantage of the fact that the filaments
lie on the $y$-$z$ plane, and noting, from visual inspection, that the
LOSs are at angles  $\sim -50^\circ$ and $\sim -20^\circ$ 
from the $z$-axis in Figs.\ \ref{fig:fil1} and
\ref{fig:fil2}, respectively, we thus have rotated the fields by
minus these amounts on this plane to orient the filaments vertically.

The resulting column density maps are shown in Fig.\ \ref{fig:vel-xi},
where the vertical axis, labeled $\xi$, is now approximately
parallel to the filament.
Superposed on these maps, we show plots of the column density-weighted
mean velocity perpendicular to the line-of-sight and approximately along
the filament (the vertical velocity in the projection of Fig.\
\ref{fig:vel-xi}). This rendering clearly shows the large-scale collapse
of the filament along its long dimension, signaled by the sharp
transition (jump) from positive velocities in the lower region to
negative velocities in the upper region.  But in addition to this global
collapse, smaller jumps in velocity associated to smaller-scale (in size
and mass) centers of collapse are also observed.  The collapsing regions
move along the filament, as they follow the large-scale filament
collapse, and so the velocity jumps around these regions are superposed
on the average infall velocity towards the larger-scale center of mass.

The velocity structure seen in Fig.\ \ref{fig:vel-xi} qualitatively
resembles that observed in a number of observational studies. In
particular, \citet{Kirk+13} report a velocity gradient of $1.4 \kms$
pc$^{-1}$ across the filament associated with the
embedded Serpens South protocluster (see their Figure 4), while
\citet{Peretto+14} report velocity gradients ranging between 0.22
and 0.63 $\kms$ pc$^{-1}$ across the filaments in SDC13. Although both the
size and velocity scales of those filaments are smaller than ours, the
velocity difference in our filaments, $\sim 5 \kms$ over length scales
of $\sim 10$ pc, also corresponds to a gradient $\sim 0.5 \kms$
pc$^{-1}$. These results are only preliminary, and a survey of
the filaments produced in our simulations should be carried out, but the
qualitative similarity between our results and the observations is
encouraging.

\subsection{Accretion onto and from the filaments}
\label{sec:accretion}

As mentioned above, the filaments are not depleted in a
free-fall time (see \S \ref{sec:ident})
as they fall onto the local collapse
centers (the clumps) within them because the filaments themselves are accreting
material from their surroundings.  This infall of material occurs at
lower densities than that of the filaments, and so the infall and the
dense filament might not be both observable with the same molecular
species.
Nevertheless, we report here the accretion structure 
in the hope that it can be ultimately compared with that of observed
filaments.

\begin{figure}[!t]
  \centerline{
  \includegraphics[width=0.50\textwidth]{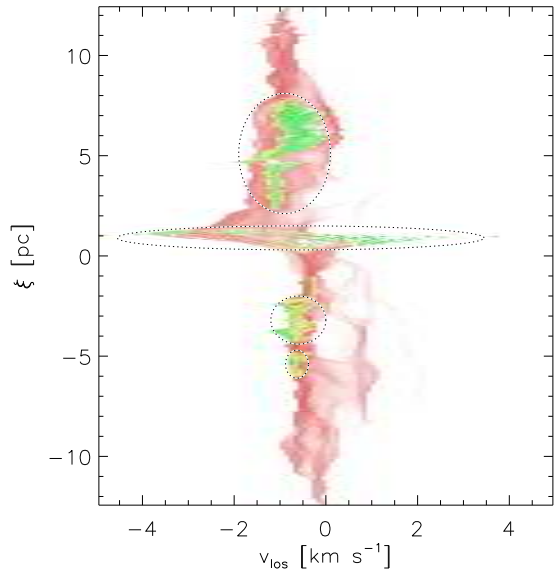}
  }
  \caption{Position-velocity map along the center of the Filament 2,
  integrated along the same direction as in fig. \ref{fig:vel-xi}
  so that the line-of-sight is approximately perpendicular to the
  filament and the $\xi$ coordinate is measured along the filament.
  The grayscale (red image in the online version of this figure)
  corresponds to gas with $n<10^3\pcc$, while contours
  (green image in the online version) corresponds to denser gas.
  Both show the column density of gas per unit velocity interval
  (on a logarithmic scale) in the range
  $3\times10^{19}\,-\,3\times10^{23}\cm^{-2}(\kms)^{-1}$.
  Denser gas breaks into separate structures, marked with
  dotted ellipses, while gas at lower density forms a continuous
  structure along the filament.
  Most of the gas in both density regimes is clearly separated in
  velocity space, with only narrow regions of overlap (yellow
  in the online version).
  }
  \label{fig:posvel}
\end{figure}

Figure \ref{fig:posvel} shows a position-velocity diagram along 
Filament 2, integrating along the same line-of-sight as in Figure
\ref{fig:vel-xi}, for gas with density both below (grayscale, red
in the online version) and above (contours, green in the online
version) $10^3\pcc$.  The velocity in this case is along the line of
sight (LOS), and denoted $\vlos$. While gas with $n < 10^3 \pcc$
forms a continuous structure along the filament, gas with $n>10^3
\pcc$ breaks into four separate structures, enclosed by dotted
ellipses in Fig.\ \ref{fig:posvel}.  One of those individual
structures is a disk surrounding a collapse region near the center of
the filament (at $\xi\approx\,0.5\pc$), distinguishable by the larger
spread in LOS velocity.  Smaller collapse centers are also noticeable
(at $\xi\approx\,-5.5,-3$, and $5\pc$). Interestingly, LOS-velocity
jumps are generally observed in both the local and the large-scale
clump-filament systems

As shown in Figure \ref{fig:fil2}, the gas falls onto the
filament in an oblique direction, and gradually becomes  a flow {\it along}
the filament as one moves towards the filament's axis.
That is, this change of direction happens as the gas density increases.
Therefore, when considering the velocity component perpendicular
to the filament ($\vlos$ in fig. \ref{fig:posvel}),
gas in different density regimes should separate in velocity space.
This is noticeable in Figure \ref{fig:posvel} as a clear separation
of the gas in these regimes
(red and green images in the online version),
with only a small superposition (visible as a yellow area in the
online version), with exception of the disk near the center, where
the nature of the gas flow is different from the rest of the filament.

The structure seen in Figure \ref{fig:posvel} is remarkably similar
to that observed by \citet{Schneider+10} in molecular line emission from
the DR21 massive filament in the Cygnus X region, which has a length
$\sim 15$ pc, very similar to that of the filaments we
analyze. \citet{Schneider+10} conclude that the kinematics of the gas is
consistent with a global gravitational collapse and inflow onto the
filament, in agreement with the dynamical state of our filaments.
Furthermore, they show, in their Figure 6, a position-velocity diagram
for several molecular transitions that closely resembles our Figure
\ref{fig:posvel}. In particular, a separation of different density
regimes in velocity space, which in our simulation corresponds to gas
falling onto the filament, is observed in their Figure 6. Moreover, the
structure in their filament also exhibits significant widenings in the
velocity coordinate, of amplitude several $\kms$, at the location of
clumps. In our simulation, these widenings are the result of the
large infall velocities that develop in the vicinity of the
clumps. These similarities strongly suggest that the filament studied by
\citet{Schneider+10} is indeed produced, like the filaments in our
simulation, by gravitational collapse.

\begin{figure}[!t]
  \centerline{ \includegraphics[width=0.50\textwidth]{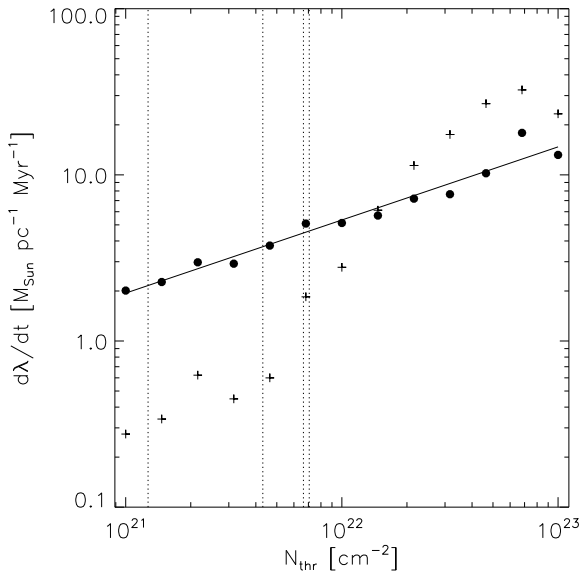} }
  \caption{ Accretion rate onto Filament 2, defined as the mass
  flux across the contour containing the column density $N_{thr}$ in the
  directions perpen\-di\-cular ({\it dots}) and parallel ({\it crosses})
  to the filament. Dotted vertical lines show column densities
  corresponding to radii $0.03, 0.1, 0.3$ and $1\pc$ ({\it right to
  left}), according to the fit to the column density profile discussed
  in \S\ref{sec:profile}. At $N > 10^{22} \psc$, the filament's radius
  cannot be derived from the column density profile fit. The structures
  at column densities larger than these are isolated clumps, rather than
  part of the fitted filament radial column density profile.  The
  perpendicular accretion may be fitted by $\dif \lambda / \dif t =
  5.35\Msun\pc^{-1}\Myr^{-1} (N/10^{22} \cm^{-2})^{0.44}$ ({\it solid
  line}).}  \label{fig:accretion}
\end{figure}

Figure \ref{fig:accretion} shows the projected accretion rate onto
Filament 2, defined as the flux across a contour defined by a given
column density value (integrated along the line-of-sight
approximately perpendicular to the filament described above).  Shown
are both the flux across longitudinal contour segments (the flux
perpendicular to the filament; filled circles) and across a contour
segment perpendicular to the filament (the flux along the filament; plus
signs).  As mentioned above, the filament accretes gas mainly in the
direction perpendicular to it.  But, since the filament itself is
falling onto the clump at the same time it is accreting, the inflow
parallel to the filament, although small, is non-negligible. For
example, at a threshold density of $3\times 10^{21} \cm^{-2}$, the flux
along the filament is a factor of $\sim 10$ smaller than the
perpendicular accretion.  In the scenario pictured above, this should
not be surprising: the filament flows along its long direction, but
keeps being replenished by its envelope.  So, at a surface defined
solely by a column density value there will be flux due to both
accretion (mainly perpendicular) onto the filament as well as flux
(parallel) corresponding to infall due to the collapse of the filament
as a whole.

The accretion rate along the filament becomes comparable to the
perpendicular one at $N_{thr} \sim 2\times10^{22}\cm^{-2}$.  This may be
understood in terms of mass conservation, since the filament is not a
hydrostatic structure, but rather, a flow feature, so that all the mass
that accretes perpendicularly to the filament must eventually flow away
in the longitudinal direction. Considering that the filament is $\sim 15
\pc$ long, its total perpendicular accretion rate is $\sim 150 \Msun
\Myr^{-1}$.

\subsection{Filament profile}
\label{sec:profile}

\subsubsection{Column density fit} \label{sec:col_dens_fit}

It has been suggested that filaments observed in molecular clouds
might have a Plummer-like profile
in the direction perpendicular to the filament's long axis.
\citet{Arzoumanian2011} explored a number of filaments observed with {\it
Herschel} and fitted a profile with a core of constant density
$\rho_c$ and radius $R_c$, and a power-law envelope of
the form

\begin{equation}
  \rho_P(R) = \frac{\rho_c}
              {\left[1+\left(R/R_c\right)^2\right]^{p/2}},
              \label{eq:plummer}
\end{equation}

\noindent
where $R$ is the distance perpendicular to the filament's long
axis, and $p$ is the (negative) logarithmic slope of the density
profile at large $R$.
The corresponding column density perpendicular to the filament is
\citep{Arzoumanian2011}

\begin{equation}
  \Sigma_P(R) = \frac{A_p \rho_c R_c}
                  {\left[1+\left(R/R_c\right)^2\right]^{(p-1)/2}},
                  \label{eq:sigma}
\end{equation}

\noindent
where $A_p = \int^\infty_{-\infty} \dif u / ( 1+u^2 )^{p/2}$.
They found that the filaments are well fitted for values of $p$ in the range
$1.5<p<2.5$, and a characteristic Gaussian FWHM of
$0.1\pc$, which corresponds to $R_c \approx 0.03\pc$.

We wish to compare the filaments in our simulation with the results of
\citet{Arzoumanian2011}. However, this is not completely
straightforward, as the column density structure around the filaments is
far from being smooth, being instead significantly clumpy. Also,
the radial column density profile is in general asymmetric around the
maximum. For example, in the lower-left panel of Fig.\ \ref{fig:fil2},
a moderate-density clump is observed at $(x,y) \approx (3,-12)\pc$,
which extends $\sim 3$ pc in the $y$-direction, while a
protrusion is observed coming off the filament at $(x,y) \approx
(-1,-5)\pc$. If a single Plummer-like profile were to be fitted to the
spatial column density distribution, these neighboring clumps
would induce strong local variations in the fitted
parameters of the column density profile namely the position of
the density maximum (the ``spine'' of the filament), the central density
$\rho_c$, the width $R_c$, and the slope $p$).

\begin{figure}[!t]
  \centerline{
  \includegraphics[width=0.50\textwidth]{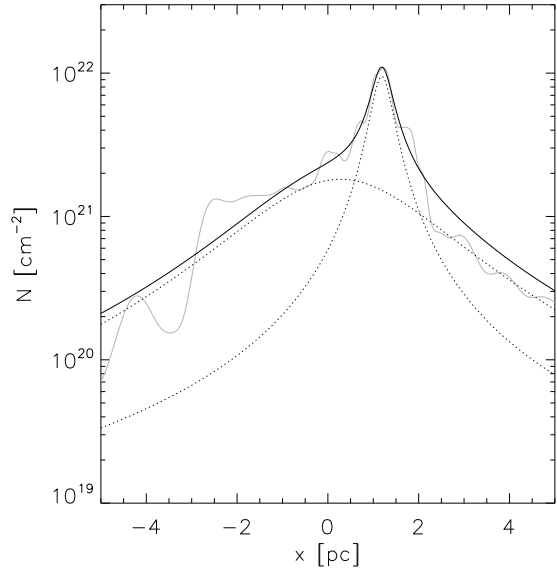}
  }
  \caption{An example of how the two-Plummer-components fit is performed
for a given radial density or column density profile. The fitted
function is the sum of two Plummer profiles, each with a different set
of parameters, so that one of them captures the position of the actual
maximum, while the other allows for an asymmetry of the physical
profile. The dotted lines show the two fitted components, while the
dark solid line shows the total fitted profile, and the gray solid line
shows the actual column density profile corresponding to the
Filament 2 at $\xi = 1.22\pc$ (see fig. \ref{fig:vel-xi}).}
  \label{fig:components}
\end{figure}

In order to minimize the impact of the irregularities, we
have chosen to use a two-step fitting procedure, as follows. First, in
order to best locate the axis of the filament, we fit
{\bf a composite profile consisting of} {\it the sum} of
two Plummer-like {\bf components}, each of the form of eq.\ (\ref{eq:sigma}),
to the column density at each position $\xi$ along the filament where the
peak column density exceeds $2\times 10^{21}\cm^{-2}$. The fit is
performed using a $\chi^2$ minimization technique. This procedure then
gives us two sets of parameters, one for each component. {\bf Figure
\ref{fig:components} illustrates this procedure.
}

In the second step, we then assume that the ``true'' position of the
filament's spine is the one given by the fitted component with the
higher central column density. We then drop all other information
from the two-component fit, and perform a new, single-component fit to
the {\it mean} radial column density profile, with the averaging
performed over both sides of this central position and along the
filament's length.

\begin{figure}[!t]
  \centerline{
  \includegraphics[width=0.50\textwidth]{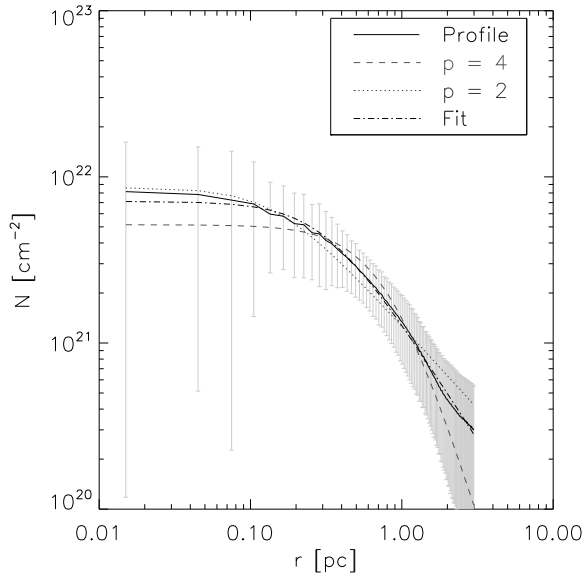}
  }
  \caption{Mean column density profile for Filament 2
  ({\it solid line}), as a function of distance perpendicular to the
  filament.
  Error bars show the spread of column density profiles for
  different values of $\xi$.
  The best fit of the form of eq. \ref{eq:sigma} ({\it dot-dashed
  line}) is shown, along with fits with $p=2,4$ ({\it dotted and
  dashed lines}, respectively) for comparison.}
  \label{fig:profile}
\end{figure}

Figure
\ref{fig:profile} shows the resulting mean profile.  The best
fit to this mean profile has $R_c=0.31\pc$,
$\rho_c/m_\eff=3.0\times10^3\pcc$,
and $p=2.4$.
This filament width is consistent with the filament data compiled
by \citet{Myers2009} for a number of nearby star-forming regions.
We also added a constant column density term
to the model in order to account for the low density diffuse gas
present in the simulation that is not observable in molecular line
emision, but
the resulting column density base is negligible
($3.8\times10^{17}\cm^{-2}$) and
the fitted profile parameters change only marginally.
In both cases, the spread in column density profiles along the
filament is much larger than the error in the fit.

\begin{figure*}[!t]
  \centerline{ \includegraphics[width=\textwidth]{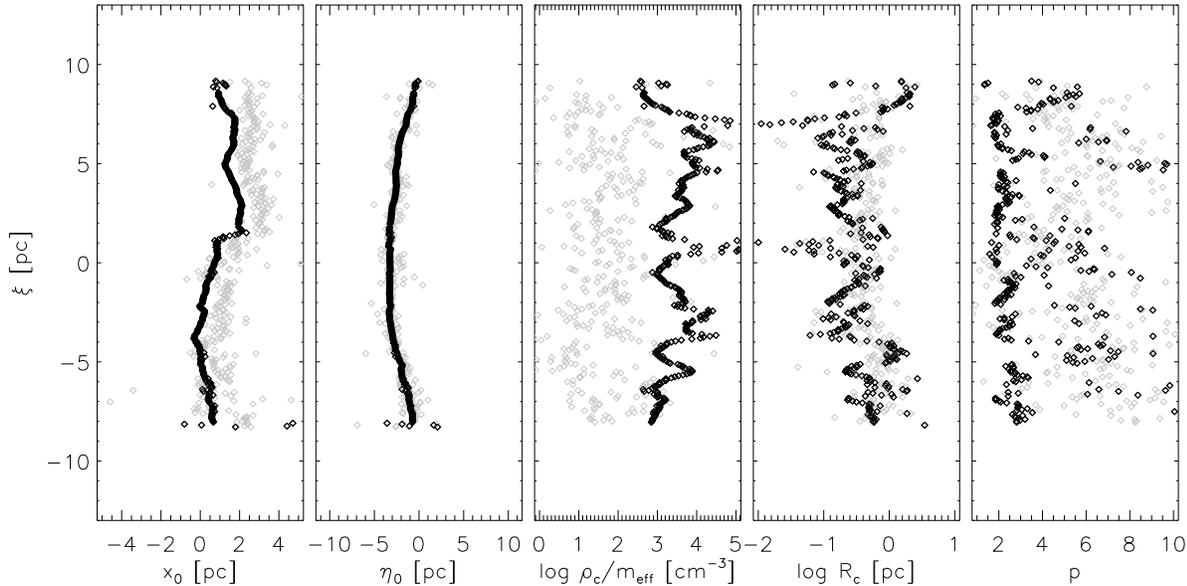} }
  \caption{Resulting parameters when a two-component profile of the form
  of eq.\ (\ref{eq:plummer}) is fitted to Filament 2. The
  parameters $x_0$ and $\eta_0$ represent the coordinates of the center
  of the profile, with $\eta$ being the direction along the
  line-of-sight (perpendicular to the plane of the figure) in
  fig.\ \ref{fig:vel-xi}.  Dark dots correspond to the main component of
  the fit, while light dots show the secondary component. The
  parameters for the secondary component are seen to fluctuate strongly,
  reflecting the random nature of the clumps in the vicinity of the main
  filament.} \label{fig:two_fils}
\end{figure*}

Figure \ref{fig:profile} also shows the fitted profile for the $p=4$
case, which corresponds to an isothermal cylinder in hydrostatic
equilibrium \citep{Ostriker1964}.
With the appropriate parameters, a $p=4$ model can be 
fitted within the large spread in column density values.
However, departures from the average profile show trends
(systematically below the mean profile in the constant core region,
for example), suggesting a bad fit.
Also, since the $p=4$ model falls more rapidly than the mean
profile, the fit is more sensitive to the constant column density
term mentioned above (in fact, this is the only case where a
non-negligible column density base was returned by the fitting
procedure, namely $3\times10^{20}\cm^{-2}$).
Regardless, a bad fit for a hydrostatic model should not be
surprising given the dynamical nature of the filament in the
simulation. In fact, Fig.\ \ref{fig:profile} also shows a
fit with $p=2$, which \citet{Arzoumanian2011} used to fit one of their
filaments. It is seen that this value of $p$ provides a much better fit
for our filament than the case $p=4$, although the best fit is obtained
for $p =2.4$.

Fitting a column density profile of the form of eq.
(\ref{eq:sigma}) to Filament 1 is less reliable than in the case of
Filament 2 since Filament 1 bends and has more nearby small scale
concentrations that skew the fit.  Nevertheless, considering only the
gas below the disk ($\xi \le 3.5\pc$) and masking out the concentration
at $(x,\xi) \approx (1,1)\pc$ (see left panel in fig. \ref{fig:vel-xi}),
the fitting procedure yielded similar central density and core radius
values ($\rho_c/m_\eff = 1.8\times10^3 \pcc$ and $R_c = 0.29\pc$),
although with a shallower profile ($p=2.0$).

\subsubsection{Volume density fit} \label{sec:vol_dens_fit}

In contrast with observations, the numerical model yields the full
spatial distribution of the gas, allowing us to explore the three
dimensional structure of the filament.  With this in mind, we repeated
the above procedure but using a profile of the form of eq.\
(\ref{eq:plummer}) to fit the volume density distribution of the
Filament 2.  In this case, the two-step fit was performed at
each $\xi$-position along the filament, this time considering the
density field in a plane perpendicular to the $\xi$ coordinate. The fit
was only performed if the maximum density on this plane, $n_{\rm 
max}$, satisfied $n_{\rm max} >10^3\pcc$. Again, the main
component is defined as that with larger central column density (the
product $\rho_c R_c$).  

Figure
\ref{fig:two_fils} shows the fitted parameters along the filament.  The
fitting procedure returns continuous positions for the center of the
main component along the filament, suggesting that it is reasonably well
defined by the fit.  But it is noteworthy that the distribution of
centers of the secondary component, although not really continuous,
tends to agglomerate to the right of the main component.  This behavior
is also present in the distribution of $\rho_c$ and $p$-values along the
filament.  We envision two possible interpretations.  One is that the
filament is really a bundle of smaller filaments, as proposed by
\citet{Hacar2013} for filaments in the Taurus region, that may be
separated in velocity space.  Another interpretation, which we find more
plausible, is that the filament is not symmetric and so, using a
symmetric form for the fit (either in real or in velocity space),
artificially yields multiple, displaced components.  This lack of
symmetry is suggested by the skewed velocities in Figure
\ref{fig:posvel}.

\begin{figure}[!t]
  \centerline{
  \includegraphics[width=0.50\textwidth]{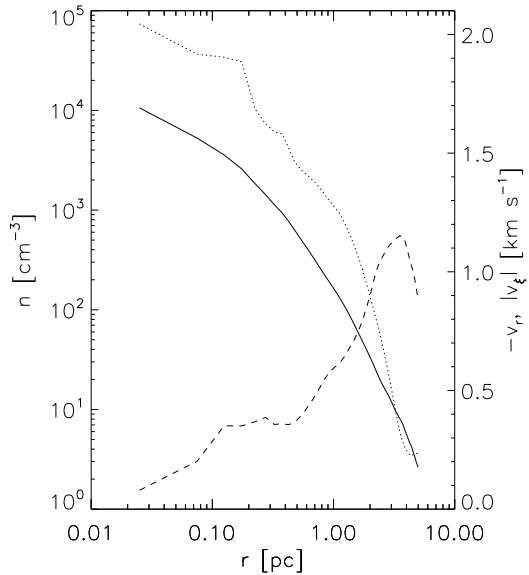}
  }
  \caption{Azimuthally averaged density ({\it solid line}),
  radial infall velocity ({\it dashed}), and longitudinal velocity
  ({\it dotted}) for the Filament 2.
  As gas is accreted onto the filament, its (radial)
  inflowing velocity gradually changes direction, so that the gas in
the centermost regions flows mainly along the
  filament.
  }
  \label{fig:rad_profile}
\end{figure}

With the center of the filament defined by the main component, we may
average to obtain a (cylindrically symmetric) density
and velocity profiles
characteristic of the filament.
Such profiles are shown in Figure \ref{fig:rad_profile}.
As the gas is accreated into the filament, the radially inflowing velocity
changes direction, turning the gas motion into a flow along the filament.
It does not appear to happen abruptly (through a shock, for
example), but in a smooth way.
This change in the flow direction is apparent also in Figures \ref{fig:fil1}
and \ref{fig:fil2}.

Following \citet{Arzoumanian2011}, the averaged radial density profile
was then fitted by a single
Plummer-like shape.  In this case, the best single-component fit has
parameters $R_c=0.11\pc$, 
$\rho_c/m_\eff=8.6\times10^3\pcc$, and $p=1.9$.  This value of $p$
implies a shallower volume density profile than the corresponding column
density one. This is, however, to be expected, as the column density,
seen in projection, increases towards the central axis of the filament
due to the combined effects of a larger volume density at the central
axis and of a longer integration path there.  Nevertheless, without the
smoothing effect provided by the line-of-sight integration, the spread
in values in the volume density profiles is much larger than in the
column density case. This sends a warning that the actual
three-dimensional structure of observed filaments may be significantly
more disordered and chaotic than what might be inferred from fits to the
column density, which constitutes a LOS-average, and are furthermore
averaged along the length of the filament.

\section{Limitations and resolution considerations} \label{sec:limit}

Our analysis is not free of caveats and limitations. First, we only
analyzed two filaments, as a proof-of-concept, and only at a single
typical scale, although filaments at different size and mass scales are
apparent in the simulation. In the future, we plan to perform a
systematic survey-like study, applying an automated filament-detection
algorithm, which should allow us to obtain significant statistics and
investigate scaling properties. 

Second, our simulations have omitted a number of physical processes,
most notably magnetic fields and stellar feedback. In particular, the
omission of stellar feedback implies that our study should be applicable
only to the early evolutionary stages of the filament-clump system. On
the other hand, magnetic fields have been suggested to have an important
role in the formation and structuring of filaments \citep[e.g.] []
{Padoan+01, HA13}, so a study analogous to the one we have presented here
should be performed including magnetic fields. Nevertheless, the fact
that our filaments compare reasonable well with observations suggests
that the magnetic field may only play a relatively minor role.

\begin{figure}[!t]
  \centerline{
  \includegraphics[width=0.50\textwidth]{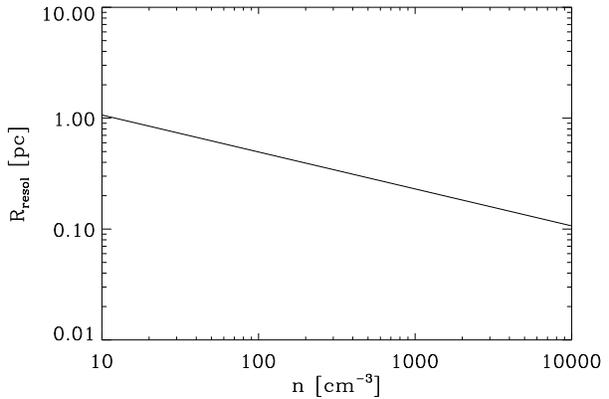}
  }
\caption{Minimum resolved size scale as a function of density, computed
as the radius of a sphere at the indicated density, and containing twice
the mass within the smoothing length of the code (i.e., the mass of 80
SPH particles, corresponding to $m=1.6 \Msun$).}
\label{fig:resol}
\end{figure}

Third, it is important to estimate the level at which our filamentary
structures are resolved. This can be done as follows. A standard
resolution criterion with SPH is that the minimally-resolved mass is
twice that contained within a smoothing length. Our simulation employed
40 neighbors to define the smoothing length. Given the SPH particle mass
of $\approx 0.02 \Msun$ (cf.\ \S\ref{sec:model}), then, the
minimally-resolved mass is $m \approx 1.6 \Msun$. To determine a
corresponding size scale at a given density, we consider the radius of a
sphere containing this mass at the specified density, given by $R_{\rm
resol} = (3 m/4 \pi \rho)^{1/3}$. Figure \ref{fig:resol} shows the
scaling of $R_{\rm resol}$ with density. This can be compared with the
size scales determined for our filaments in \S\ref{sec:profile}.
There, we found that the {\it column} density profiles were well fitted
by $R_c \approx 0.3$ pc and $\rho_c/{m_{\eff}} \sim $2--3 $\times 10^ 3
\pcc$. By comparison, we see from Fig.\ \ref{fig:resol} that, at $n=2.5
\times 10^3 \pcc$, $R_{\rm resol} \approx 0.17$ pc, so that the fitted
central scale is roughly twice the minimum resolved scale, and we can
consider that this fit is safely resolved.

On the other hand, also in \S\ref{sec:model}, we found that the {\it
volume} density profile was fitted with $R_c \approx 0.11$ pc and
$\rho_c/{m_{\eff}} \approx 8.6 \times 10^ 3 \pcc$. For comparison, we see
from Fig.\ \ref{fig:resol} that, at $n= 8.6 \times 10^3 \pcc$, $R_{\rm
resol} \approx 0.11$ pc, so in this case the fit is only marginally
resolved. Thus, we cannot rule out the possibility that the
characteristic width of our filaments may be limited by resolution.

Nevertheless, despite the relative simplicity and resolution limitations
of the simulations presented here, the fact that a simulation originally
designed to study a different phenomenon (the evolution of the star
formation activity before feedback dominates) yields filaments similar to
observed ones suggests that the dominant physical processes involved in
the filament dynamics are present in the simulation.

\section{Summary and Conclusions}
\label{sec:summary}

We performed SPH simulations of the formation of a molecular cloud
from a convergent flow of diffuse gas.
As in previous works, the resulting cloud is highly dynamic
and, due to hydrodynamic and thermal instabilities, becoming turbulent and
rapidly fragmenting into substructure.
Nevertheless,
the ``turbulent'' motions are unable to stabilize the cloud and soon
its dynamics becomes dominated by gravity, and the cloud begins to
collapse. However, because the cloud continues to accrete, its mass
continues to grow, and soon it contains a large number of Jeans masses
\citep{VazquezSemadeni2007}. This naturally leads to the formation of
filaments, since the multi-Jeans mass flow behaves similarly to a
pressureless flow, which is known to collapse along its shortest
dimension first \citep{Lin+65}, so that the initially planar cloud
collapses onto filaments, which then collapse into clumps.

The filaments are not equilibrium structures at all, and instead are
highly dynamical. They can be considered intermediate steps of the
cloud's collapse, constituting channels through which the gas is
funneled from the extended cloud onto the clumps.  These filaments are
the loci of colliding streams on the parent planar structures, where
the flow gets reoriented and directed towards its final destination --
the clumps within the filaments, which have conditions appropriate for
star formation.

It is important to remark that the filaments developing in our
simulation arise self-consistently from the fragmentation of the cloud
initially formed by the convergence of warm gas streams by the combined
action of various instabilities \citep{Heitsch+05}. In particular,
thermal instability produces small, cold clumps that grow by the
accretion of warm gas as well as by merging with other clumps
\citep{Banerjee+09}, until they finally become gravitationally unstable
and begin to collapse. Thus, our simulation differs fundamentally from
simulations of {\it Fourier-generated} turbulence in the multi-phase
medium, in which the external velocity field is constructed in Fourier
space and is applied everywhere in space \citep[e.g.,] [] {VGS00, VS+03,
Gazol+05, Gazol+09, Goodwin+04, Goodwin+06, GK10, GK13, Bate09,
Walch+12}. The latter method allows precise control of the driving
scale, the amplitude, and nature (ratio of solenoidal to compressible
energy) of the applied force, albeit at the expense that the driving
is applied as a body force rather than as a surface stress, as
corresponds to the kind of energy sources present in the ISM, such as
outflows and expanding shells. Instead, in our simulation, the
characteristic (or ``energy-containing'') scale of the turbulence, as
well as the amplitude and ratio of solenoidal to compressible, develop
self-consistently from the instabilities in the compressed layer, and we
have no control over these parameters.

Under these conditions, in our simulation, the gravitational collapse
in the individual centers (modeled with sink particles), is not
instantaneous but starts $\sim 20\Myr$ into the simulation and continues
well after the filaments finish to accrete onto the clumps. Moreover,
the filaments themselves are falling into the largest-scale potential
well, generating new structures that eventually merge into each other,
forming a larger-scale hub-filament system towards the end of the
simulation.

The simulation presented here formed two prominent filaments, each with
$\sim 600\Msun$ out to radii $\sim 0.3 \pc$, and lengths $\sim 15\pc$,
implying linear mass densities $\lambda \sim 40 \Msun \pc^{-1}$.  We
furthermore studied the radial density and column density profiles of
such filaments,
finding that the column density profile is reasonably well
fitted by a Plummer-like profile, whose best-fit parameters are $R_c =
0.31 \pc$ and $p = 2.4$. In turn, the three-dimensional radial density
distribution was best fitted with $R_c = 0.1 \pc$ and $p\approx
1.9$, although this fit was less certain because in the
three-dimensional measurement no averaging along the LOS is made.

These values are qualitatively consistent with observational
results. For example, \citet{Andre2010} find that their sample of
filaments in the Aquila region have linear mass densities $\lambda > 15
\Msun \pc^{-1}$, while \citet{Arzoumanian2011} find that their filaments
typically have Plummer-type column density radial profiles with $R_c
\sim 0.1 \pc$ and $1.5 < p < 2.5$. These authors speculate that their
filaments may form as the result of the dissipation of large-scale
turbulence, because their characteristic thickness is similar to that of
the sonic scale, below which the ``turbulence'' becomes
subsonic. However, in our case, the filaments are not a result of
turbulence nor their thickness is the result of having dissipated the
turbulence. Instead, they are an intermediate stage of the gravitational
collapse, funneling gas to the local collapse centers -- the clumps --
and their thickness appears to be the result of the force
balance between the radial pressure gradient in the filament and the
combined thermal plus ram pressure of the gas falling onto the filament,
the whole system maintained in a stationary state thanks to the
``drainage'' and reorientation of the infalling material towards the
trough of the gravitational potential well, located at the clumps
accreting from the filaments. We plan to will explore this scenario in a
future study.

We also measured the accretion rates perpendicular (onto) and parallel
(along) the filament. The rates are not unique, but instead depend on
the radial position along the filament. Because of the column density
profile, this translates into a column density dependence of the
accretion rates. We found that, as the column density varies from
$10^{21}$ to $10^{23} \psc$, the accretion rate per unit length
perpendicular to the filament (i.e., onto it) ranges between 2 and 20
$\Msun \Myr^{-1} \pc^{-1}$, while the linear accretion density from the
filament onto the clump ranges between 0.3 and 30 $\Msun \Myr^{-1}
\pc^{-1}$. Considering that the filament is $\sim 15 \pc$ long, its
total perpendicular accretion rate is $\sim 150 \Msun \Myr^{-1}$.

Our measured accretion rates, which peak at 2--3$\,
\times 10 \Msun \pc^{-1} \Myr^{-1}$, are at least half an order of
magnitude lower than those inferred by \citet{Schisano+13}, although
their estimate was an indirect one, assuming that the characteristic
timescale for the process of moving the material along the filament is
equal to that of the lifetime of the protostellar objects being formed
in the clumps, $\sim 10^4$ yr. This is a rather uncertain assumption,
which may overestimate the accretion rate estimate. On the other hand,
our total measured accretion rate of $\sim 150 \Msun \Myr^{-1}$  matches
very well that inferred by \citet{Kirk+13} ($\sim 130 \Msun \Myr^{-1}$).
However, the filament they studied was at a much smaller
(sub-pc) scale, and therefore the comparison may not be conclusive.

We conclude from our results that the process of hierarchical
(i.e., multi-scale) and chaotic (i.e., disordered) gravitational
fragmentation produces filaments that have physical conditions and
accretion rates that are qualitatively similar to those found in recent
observational studies, suggesting that this mechanism may indeed be
responsible for their formation. However, more systematic studies need
to be performed where the mass and size scales of the filaments in both
the observations and simulations are better matched, in order to
determine whether in this case their radial column density profiles and
accretion rates match at the quantitative level. We plan to perform such
a study in the near future.

\acknowledgments

We thankfully acknowledge N. Schneider for useful comments on this
work, and two anonymous referees, whose reports helped
in significantly improving the clarity of the paper.  This work has
received financial support from UNAM-DGAPA PAPIIT grant IN111313 to
GCG and CONACYT grant 102488 to EVS. The numerical simulation was
performed on the cluster at our Center acquired through the latter
grant.


\begin{thebibliography}{}

\bibitem[\protect\citeauthoryear{Andr\'{e} et~al.,}{Andr\'{e}
  et~al.}{2010}]{Andre2010}
Andr\'{e} P.  et~al., 2010, A\&A, 518, L102

\bibitem[\protect\citeauthoryear{Arzoumanian et~al.,}{Arzoumanian
  et~al.}{2011}]{Arzoumanian2011}
Arzoumanian D.  et~al., 2011, A\&A, 529, L6

\bibitem[Audit \& Hennebelle(2005)]{AH05} Audit, E., \& Hennebelle, P.\ 2005, \aap, 433, 1 

\bibitem[Bally et al.(1987)]{Bally+87} Bally, J., Lanber, W.~D., 
Stark, A.~A., \& Wilson, R.~W.\ 1987, \apjl, 312, L45 

\bibitem[Balsara et al.(2001)]{Balsara+01} Balsara, D., 
Ward-Thompson, D., \& Crutcher, R.~M.\ 2001, \mnras, 327, 715 

\bibitem[Banerjee et al.(2009)]{Banerjee+09} Banerjee, R., 
V{\'a}zquez-Semadeni, E., Hennebelle, P., 
\& Klessen, R.~S.\ 2009, \mnras, 398, 1082 

\bibitem[Bate(2009)]{Bate09} Bate, M.~R.\ 2009, \mnras, 397,  232 

\bibitem[Bate et al.(1995)]{Bate+95} Bate, M.~R., Bonnell, 
I.~A., \& Price, N.~M.\ 1995, \mnras, 277, 362 

\bibitem[Battersby et al.(2014)]{Battersby+14} Battersby, C., 
Ginsburg, A., Bally, J., et al.\ 2014, \apj, 787, 113

\bibitem[Bond et al.(1996)]{Bond+96} Bond, J.~R., Kofman, L., 
\& Pogosyan, D.\ 1996, \nat, 380, 603 

\bibitem[Burkert \& Hartmann(2004)]{BH04} Burkert, A., \& Hartmann, L.\
2004, \apj, 616, 288 

\bibitem[Cantalupo et al.(2014)]{Cantalupo+14} Cantalupo, S., 
Arrigoni-Battaia, F., Prochaska, J.~X., Hennawi, J.~F., 
\& Madau, P.\ 2014, \nat, 506, 63 

\bibitem[Carroll-Nellenback et al.(2014)]{Carroll+14} 
Carroll-Nellenback, J., Frank, A., \& Heitsch, F.\ 2013, arXiv:1304.1367 

\bibitem[Clark \& Bonnell(2005)]{CB05} Clark, P.~C., \& Bonnell, I.~A.\
2005, \mnras, 361, 2 

\bibitem[Col{\'{\i}}n et al.(2013)]{Colin+13} Col{\'{\i}}n, P., 
V{\'a}zquez-Semadeni, E., \& G{\'o}mez, G.~C.\ 2013, \mnras, 435, 1701

\bibitem[Csengeri et al.(2011)]{Csengeri+11} Csengeri, T., Bontemps, S.,
Schneider, N., Motte, F., \& Dib, S.\ 2011, \aap, 527, A135 

\bibitem[Curry(2000)]{Curry00} Curry, C.~L.\ 2000, \apj, 541, 
831 

\bibitem[de Avillez \& Breitschwerdt(2004)]{AB04} de Avillez, M.~A., \&
Breitschwerdt, D.\ 2004, \aap, 425, 899 

\bibitem[Dickey et al.(1977)]{Dickey+77} Dickey, J.~M., Salpeter, 
E.~E., \& Terzian, Y.\ 1977, \apjl, 211, L77 

\bibitem[Dobbs et al.(2011)]{Dobbs+11} Dobbs, C.~L., Burkert, 
A., \& Pringle, J.~E.\ 2011, \mnras, 413, 2935

\bibitem[Elmegreen(1987)]{Elm87} Elmegreen, B.~G.\ 1987, \apj, 312, 626 

\bibitem[Federrath et al.(2010)]{Federrath+10} Federrath, C., 
Banerjee, R., Clark, P.~C., \& Klessen, R.~S.\ 2010, \apj, 713, 269 

\bibitem[Federrath \& Klessen(2012)]{FK12} Federrath, C., \& Klessen,
R.~S.\ 2012, \apj, 761, 156 

\bibitem[Federrath \& Klessen(2013)]{FK13} Federrath, C., \& Klessen,
R.~S.\ 2013, \apj, 763, 51 

\bibitem[Feitzinger et al.(1987)]{Feitzinger+87} Feitzinger, J.~V., 
Perschke, M., Haynes, R.~F., Klein, U., 
\& Wielebinski, R.\ 1987, Vistas in Astronomy, 30, 243 

\bibitem[Field et al.(2008)]{Field+08} Field, G.~B., Blackman, 
E.~G., \& Keto, E.~R.\ 2008, \mnras, 385, 181 

\bibitem[Field et al.(1969)]{Field+69} Field, G.~B., Goldsmith, 
D.~W., \& Habing, H.~J.\ 1969, \apjl, 155, L149 

\bibitem[Galv{\'a}n-Madrid et al.(2009)]{Galvan+09} 
Galv{\'a}n-Madrid, R., Keto, E., Zhang, Q., et al.\ 2009, \apj, 706, 1036 

\bibitem[Gazol \& Kim(2010)]{GK10} Gazol, A., \& Kim, J.\ 2010, \apj, 723, 482 

\bibitem[Gazol \& Kim(2013)]{GK13} Gazol, A., \& Kim, J.\ 2013, \apj, 765, 49 

\bibitem[Gazol et al.(2009)]{Gazol+09} Gazol, A., Luis, L., 
\& Kim, J.\ 2009, \apj, 693, 656 

\bibitem[Gazol et al.(2005)]{Gazol+05} Gazol, A., 
V{\'a}zquez-Semadeni, E., \& Kim, J.\ 2005, \apj, 630, 911 

\bibitem[Gazol et al.(2001)]{Gazol+01} Gazol, A., 
V{\'a}zquez-Semadeni, E., S{\'a}nchez-Salcedo, F.~J., 
\& Scalo, J.\ 2001, \apjl, 557, L121 

\bibitem[Gehman et al.(1996)]{Gehman+96} Gehman, C.~S., Adams, 
F.~C., \& Watkins, R.\ 1996, \apj, 472, 673 

\bibitem[Goldsmith et al.(2008)]{Goldsmith+08} Goldsmith, P.~F., 
Heyer, M., Narayanan, G., et al.\ 2008, \apj, 680, 428 

\bibitem[\protect\citeauthoryear{G\'{o}mez, V\'{a}zquez-Semadeni, Shadmehri \&
  Ballesteros-Paredes}{G\'{o}mez et~al.}{2007}]{Gomez2007}
G\'{o}mez G.~C.,  V\'{a}zquez-Semadeni E.,  Shadmehri M.,
  Ballesteros-Paredes J.,  2007, ApJ, 669, 1042

\bibitem[Goodwin et al.(2004)]{Goodwin+04} Goodwin, S.~P., Whitworth,
A.~P., \& Ward-Thompson, D.\ 2004, \aap, 423, 169 

\bibitem[Goodwin et al.(2006)]{Goodwin+06} Goodwin, S.~P., Whitworth,
A.~P., \& Ward-Thompson, D.\ 2006, \aap, 452, 487 

\bibitem[Gutermuth et al.(2008)]{Gutermuth+08} Gutermuth, R.~A., 
Bourke, T.~L., Allen, L.~E., et al.\ 2008, \apjl, 673, L151 

\bibitem[Hartmann et al.(2001)]{Hartmann+01} Hartmann, L., 
Ballesteros-Paredes, J., \& Bergin, E.~A.\ 2001, \apj, 562, 852 

\bibitem[Hartmann \& Burkert(2007)]{HB07} Hartmann, L., \& Burkert, A.\
2007, \apj, 654, 988 

\bibitem[Hacar et al.(2013)]{Hacar2013} Hacar, A., 
Tafalla, M., Kauffmann, J., \& Kov\'acs, A.\ 2013, \aap, 554, 55 

\bibitem[Heiles \& Troland(2003)]{HT03} Heiles, C., \& Troland, T.~H.\
2003, \apj, 586, 1067 

\bibitem[Heitsch et al.(2009)]{Heitsch+09} Heitsch, F., 
Ballesteros-Paredes, J., \& Hartmann, L.\ 2009, \apj, 704, 1735 

\bibitem[Heitsch et al.(2005)]{Heitsch+05} Heitsch, F., Burkert, 
A., Hartmann, L.~W., Slyz, A.~D., 
\& Devriendt, J.~E.~G.\ 2005, \apjl, 633, L113 

\bibitem[Heitsch \& Hartmann(2008)]{HH08} Heitsch, F., \& Hartmann, L.\
2008, \apj, 689, 290 

\bibitem[Heitsch et al.(2008b)]{Heitsch+08b} Heitsch, F., Hartmann, 
L.~W., \& Burkert, A.\ 2008, \apj, 683, 786 

\bibitem[Heitsch et al.(2008a)]{Heitsch+08a} Heitsch, F., Hartmann, 
L.~W., Slyz, A.~D., Devriendt, J.~E.~G., 
\& Burkert, A.\ 2008, \apj, 674, 316 

\bibitem[Heitsch et al.(2006)]{Heitsch+06} Heitsch, F., Slyz, 
A.~D., Devriendt, J.~E.~G., Hartmann, L.~W., 
\& Burkert, A.\ 2006, \apj, 648, 1052 

\bibitem[Hennebelle \& Andr{\'e}(2013)]{HA13} Hennebelle, P., \&
Andr{\'e}, P.\ 2013, \aap, 560, A68 

\bibitem[Hennebelle \& Chabrier(2011)]{HC11} Hennebelle, P., \&
Chabrier, G. \ 2011, ApJ, 743, L29 

\bibitem[Hennebelle \& P{\'e}rault(1999)]{HP99} Hennebelle, P., \&
P{\'e}rault, M.\ 1999, \aap, 351, 309 

\bibitem[Henning et al.(2010)]{Henning+10} Henning, T., Linz, H.,
Krause, O., et al.\ 2010, \aap, 518, L95 

\bibitem[Hoyle(1953)]{Hoyle53} Hoyle, F.\ 1953, \apj, 118, 513

\bibitem[Hunter et al.(1986)]{Hunter+86} Hunter, J.~H., Jr., 
Sandford, M.~T., II, Whitaker, R.~W., \& Klein, R.~I.\ 1986, \apj, 305, 309 

\bibitem[Jappsen et al.(2005)]{Jappsen+05} Jappsen, A.-K., Klessen,
R.~S., Larson, R.~B., Li, Y., \& Mac Low, M.-M.\ 2005, \aap, 435, 611 

\bibitem[Jenkins \& Tripp(2011)]{JT11} Jenkins, E.~B., \& Tripp, T.~M.\
2011, \apj, 734, 65 

\bibitem[Juvela et al.(2009)]{Juvela+09} Juvela, M., Pelkonen, V.-M., \&
Porceddu, S.\ 2009, \aap, 505, 663 

\bibitem[Kalberla et al.(1985)]{Kalberla+85} Kalberla, P.~M.~W.,
Schwarz, U.~J., \& Goss, W.~M.\ 1985, \aap, 144, 27 

\bibitem[Kim et al.(2002)]{Kim+02} Kim, W.-T., Ostriker, 
E.~C., \& Stone, J.~M.\ 2002, \apj, 581, 1080 

\bibitem[Kirk et al.(2013)]{Kirk+13} Kirk, H., Myers, P.~C., 
Bourke, T.~L., et al.\ 2013, \apj, 766, 115 

\bibitem[Klein \& Woods(1998)]{KW98} Klein, R.~I., \& Woods, D.~T.\
1998, \apj, 497, 777 

\bibitem[Klessen et al.(2000)]{KHM00} Klessen, R.~S., 
Heitsch, F., \& Mac Low, M.-M.\ 2000, \apj, 535, 887 

\bibitem[Klessen \& Hennebelle(2010)]{KH10} Klessen, R.~S., \&
Hennebelle, P.\ 2010, \aap, 520, A17  

\bibitem[Kolb \& Turner(1990)]{KT90} Kolb, E.~W., \& Turner, M.~S.\
1990, Front.~Phys., Vol.~69

\bibitem[Koyama \& Inutsuka(2000)]{KI00} Koyama, H., \& Inutsuka, S.-I.\
2000, \apj, 532, 980 

\bibitem[Koyama \& Inutsuka(2002)]{KI02} Koyama, H., \& Inutsuka, S.-i.\ 2002, \apjl, 564, L97 

\bibitem[Kritsuk et al.(2007)]{Kritsuk+07} Kritsuk, A.~G., Norman, 
M.~L., Padoan, P., \& Wagner, R.\ 2007, \apj, 665, 416 

\bibitem[Krumholz \& McKee(2005)]{KM05} Krumholz, M. R., \& McKee,
C. F. \ 2005, ApJ, 630, 250

\bibitem[Larson(1985)]{Larson85} Larson, R.~B.\ 1985, \mnras, 
214, 379 

\bibitem[Lin et al.(1965)]{Lin+65} Lin, C.~C., Mestel, L., 
\& Shu, F.~H.\ 1965, \apj, 142, 1431 

\bibitem[McKee \& Ostriker(1977)]{MO77} McKee, C.~F., \& Ostriker,
J.~P.\ 1977, \apj, 218, 148 

\bibitem[Men'shchikov et al.(2010)]{Menshchikov+10} Men'shchikov, A.,
Andr{\'e}, P., Didelon, P., et al.\ 2010, \aap, 518, L103 

\bibitem[Micic et al.(2013)]{Micic+13} Micic, M., Glover, 
S.~C.~O., Banerjee, R., \& Klessen, R.~S.\ 2013, \mnras, 432, 626 

\bibitem[Miyama et al.(1987)]{Miyama+87} Miyama, S.~M., Narita, 
S., \& Hayashi, C.\ 1987, Progress of Theoretical Physics, 78, 1051 

\bibitem[Molinari et al.(2010)]{Molinari+10} Molinari, S., Swinyard, B.,
Bally, J., et al.\ 2010, \aap, 518, L100 

\bibitem[\protect\citeauthoryear{Myers}{Myers}{2009}]{Myers2009}
Myers P.~C.,  2009, ApJ, 700, 1609

\bibitem[Nagai et al.(1998)]{Nagai+98} Nagai, T., Inutsuka, 
S.-I., \& Miyama, S.~M.\ 1998, \apj, 506, 306 

\bibitem[Naranjo-Romero et al.(2012)]{Naranjo+12} Naranjo-Romero, 
R., Zapata, L.~A., V{\'a}zquez-Semadeni, E., et al.\ 2012, \apj, 757, 58 

\bibitem[\protect\citeauthoryear{Ostriker}{Ostriker}{1964}]{Ostriker1964}
Ostriker J.,  1964, ApJ, 140, 1056

\bibitem[Padoan(1995)]{Padoan95} Padoan, P.\ 1995, \mnras, 277, 377 

\bibitem[Padoan et al.(2001)]{Padoan+01} Padoan, P., Juvela, M., 
Goodman, A.~A., \& Nordlund, {\AA}.\ 2001, \apj, 553, 227 

\bibitem[Padoan \& Nordlund(1999)]{PN99} Padoan, P., \& Nordlund,
{\AA}.\ 1999, \apj, 526, 279 

\bibitem[Padoan \& Nordlund(2002)]{PN02} Padoan, P., \& Nordlund,
{\AA}.\ 2002, \apj, 576, 870 

\bibitem[Padoan \& Nordlund(2011)]{PN11} Padoan, P., \& Nordlund, A., \
2011, ApJ, 730, 40 

\bibitem[Padoan et al.(2007)]{Padoan+07} Padoan, P., Nordlund, 
{\AA}., Kritsuk, A.~G., Norman, M.~L., \& Li, P.~S.\ 2007, \apj, 661, 972 

\bibitem[Pavlovski et al.(2002)]{Pavlovski+02} Pavlovski, G., Smith, 
M.~D., Mac Low, M.-M., \& Rosen, A.\ 2002, \mnras, 337, 477 

\bibitem[Peretto et al.(2014)]{Peretto+14} Peretto, N., Fuller, G.~A.,
Andr{\'e}, P., et al.\ 2014, \aap, 561, A83 

\bibitem[Pon et al.(2012)]{Pon+12} Pon, A., Toal{\'a}, J.~A., 
Johnstone, D., et al.\ 2012, \apj, 756, 145 

\bibitem[Schisano et al.(2013)]{Schisano+13} Schisano, E. 2013, submitted

\bibitem[Schneider et al.(2010)]{Schneider+10} Schneider, N., Csengeri,
T., Bontemps, S., et al.\ 2010, \aap, 520, A49 

\bibitem[Schneider et al.(2012)]{Schneider+12} Schneider, N., Csengeri,
T., Hennemann, M., et al.\ 2012, \aap, 540, L11 

\bibitem[Smith et al.(2012)]{Smith+12} Smith, R.~J., Shetty, R., 
Stutz, A.~M., \& Klessen, R.~S.\ 2012, \apj, 750, 64 

\bibitem[Spitzer \& Fitzpatrick(1995)]{SF95} Spitzer, L., Jr., \&
Fitzpatrick, E.~L.\ 1995, \apj, 445, 196 

\bibitem[Tasker \& Bryan(2008)]{TB08} Tasker, E.~J., \& Bryan, G.~L.\
2008, \apj, 673, 810 

\bibitem[Vazquez-Semadeni(2012a)]{VS12} Vazquez-Semadeni, E.\ 
2012, in ``Magnetic Fields in Diffuse Media'', eds.\ E.\ de Gouveia dal
Pino, A. Lazarian, in press (arXiv:1208.4132)

\bibitem[V{\'a}zquez-Semadeni(2012)]{VS12b} 
V{\'a}zquez-Semadeni, E.\ 2012, EAS Publications Series, 56, 39 

\bibitem[V{\'a}zquez-Semadeni et al.(2003b)]{VS+03b} 
V{\'a}zquez-Semadeni, E., Ballesteros-Paredes, J., 
\& Klessen, R.~S.\ 2003, \apjl, 585, L131 

\bibitem[V{\'a}zquez-Semadeni et al.(2011)]{VS+11} 
V{\'a}zquez-Semadeni, E., Banerjee, R., G{\'o}mez, G.~C., et al.\ 2011, 
\mnras, 414, 2511 

\bibitem[V\'{a}zquez-Semadeni et~al.(2010)]{Vazquez-Semadeni2010}
V\'{a}zquez-Semadeni E.,  Col\'{\i}n P.,  G\'{o}mez G.~C.,  Ballesteros-Paredes
  J.,    Watson A.~W.,  2010, ApJ, 715, 1302

\bibitem[V{\'a}zquez-Semadeni et al.(2003a)]{VS+03} 
V{\'a}zquez-Semadeni, E., Gazol, A., Passot, T., 
\& et al.\ 2003, Turbulence and Magnetic Fields in Astrophysics, 614, 213 

\bibitem[V{\'a}zquez-Semadeni et al.(2000)]{VGS00} 
V{\'a}zquez-Semadeni, E., Gazol, A., \& Scalo, J.\ 2000, \apj, 540, 271 

\bibitem[V\'{a}zquez-Semadeni et~al.(2007)]{VazquezSemadeni2007}
V\'{a}zquez-Semadeni E.,  G\'{o}mez G.~C.,  Jappsen A.-K.,  Ballesteros-Paredes
  J.,  Gonz\'{a}lez R.~F.,    Klessen R.~S.,  2007, ApJ, 657, 870

\bibitem[V\'{a}zquez-Semadeni et~al.(2009)]{Vazquez-Semadeni2009}
V\'{a}zquez-Semadeni E.,  G\'{o}mez G.~C.,  Jappsen A.-K.,  Ballesteros-Paredes
  J.,    Klessen R.~S.,  2009, ApJ, 707, 1023

\bibitem[V{\'a}zquez-Semadeni et al.(2008)]{VS+08} 
V{\'a}zquez-Semadeni, E., Gonz{\'a}lez, R.~F., Ballesteros-Paredes, J., 
Gazol, A., \& Kim, J.\ 2008, \mnras, 390, 769 

\bibitem[V{\'a}zquez-Semadeni et al.(2005)]{VS+05} 
V{\'a}zquez-Semadeni, E., Kim, J., Shadmehri, M., 
\& Ballesteros-Paredes, J.\ 2005, \apj, 618, 344 

\bibitem[V{\'a}zquez-Semadeni et~al.(2006)]{VS+06} 
V{\'a}zquez-Semadeni, E., Ryu, D., Passot, T., Gonz{\'a}lez, R.~F., 
\& Gazol, A.\ 2006, \apj, 643, 245 

\bibitem[\protect\citeauthoryear{Toal\'{a}, V\'{a}zquez-Semadeni \&
  G\'{o}mez}{Toal\'{a} et~al.}{2012}]{Toala}
Toal\'{a} J.~A.,  V\'{a}zquez-Semadeni E.,    G\'{o}mez G.~C.,  2012, ApJ, 744,
  190

\bibitem[Vishniac(1994)]{Vishniac94} Vishniac, E.~T.\ 1994, \apj, 
428, 186 

\bibitem[Wada \& Norman(2007)]{WN07} Wada, K., \& Norman, C.~A.\ 2007,
\apj, 660, 276 

\bibitem[Walch et al.(2012)]{Walch+12} Walch, S., Whitworth, 
A.~P., \& Girichidis, P.\ 2012, \mnras, 419, 760 

\bibitem[Walder \& Folini(2000)]{WF00} Walder, R., \& Folini, D.\ 2000,
\apss, 274, 343 

\end{thebibliography}


\end{document}